\documentclass[aps,prd,twocolumn,preprintnumbers,groupedaddress,nofootinbib,amssymb,notitlepage,eqsecnum]{revtex4-2}
\usepackage{bm}
\usepackage{amsmath,amsthm,amssymb}
\usepackage{amsfonts}
\usepackage{hyperref}
\allowdisplaybreaks[1]

\newcommand{\nn}{\nonumber}
\newcommand{\rd}{{\rm d}}
\newcommand{\be}{\begin{equation}}
\newcommand{\ee}{\end{equation}}
\newcommand{\ba}{\begin{eqnarray}}
\newcommand{\ea}{\end{eqnarray}}

\begin{document}

\preprint{YITP-24-65, IPMU24-0024, WUCG-24-05}

\title{Revisiting linear stability of black hole odd-parity perturbations \\
in Einstein-Aether gravity}

\author{Shinji Mukohyama$^{a, b}$}
\email{shinji.mukohyama@yukawa.kyoto-u.ac.jp} 

\author{Shinji Tsujikawa$^{c}$}
\email{tsujikawa@waseda.jp} 
 
\author{Anzhong Wang$^{d}$}
\email{anzhong$\_$wang@baylor.edu}

\affiliation{$^{a}$Center for Gravitational Physics, Yukawa Institute for Theoretical Physics, 
Kyoto University, 606-8502, Kyoto, Japan \\
$^{b}$Kavli Institute for the Physics and Mathematics of the Universe (WPI),
The University of Tokyo Institutes for Advanced Study,
The University of Tokyo, Kashiwa, Chiba 277-8583, Japan\\
$^{c}$Department of Physics, Waseda University, 3-4-1 Okubo, 
Shinjuku, Tokyo 169-8555, Japan\\
 $^{d}$GCAP-CASPER, Physics Department, Baylor
University, Waco, TX 76798-7316, USA}
 
\date{\today}

\begin{abstract}

In Einstein-Aether gravity, we revisit the issue of linear stabilities of black holes against odd-parity perturbations on a static and spherically symmetric background. In this theory, superluminal propagation is allowed and there is a preferred timelike direction along the unit Aether vector field. If we choose the usual spherically symmetric background coordinates with respect to the Killing time $t$ and the areal radius $r$, it may not be appropriate for unambiguously determining the black hole stability because the constant $t$ hypersurfaces are not necessarily always spacelike. Unlike past related works of black hole perturbations, we choose an Aether-orthogonal frame in which the timelike Aether field is orthogonal to spacelike hypersurfaces over the whole background spacetime. In the short wavelength limit, we show that no-ghost conditions as well as radial and angular propagation speeds coincide with those of vector and tensor perturbations on the Minkowski background. Thus, the odd-parity linear stability of black holes for large radial and angular momentum modes is solely determined by constant coefficients of the Aether derivative couplings.

\end{abstract}


\maketitle

\section{Introduction}
\label{Intro}

General relativity (GR) is a fundamental pillar of modern 
physics for describing the gravitational interaction. 
GR enjoys the invariance under transformations 
in the Lorentz group. 
From the perspective of quantum gravity and high-energy theories, however, 
there are some indications that Lorentz invariance may not be 
an exact symmetry at all energies \cite{Kostelecky:1988zi,Gambini:1998it,Douglas:2001ba,Carroll:2001ws,Amelino-Camelia:2008aez}.
Lorentz violation at high energies may allow for the possibility of 
regularizing field theories, while recovering Lorentz symmetry
at low energies \cite{Chadha:1982qq}. 
Although broken Lorentz invariance 
for the standard model matter fields is highly constrained from 
numerous experiments, the bounds on Lorentz violation 
in the gravitational sector are not so stringent 
yet \cite{Mattingly:2005re,Kostelecky:2008ts,Will:2014kxa}.

To accommodate broken Lorentz invariance for the gravitational 
fields without losing the covariant property of GR, there is a way 
of introducing a unit timelike vector field $u^{\mu}$ satisfying the 
relation $u^{\mu} u_{\mu}=-1$. This is known as Einstein-Aether 
theory \cite{Jacobson:2000xp}, in which a preferred threading with respect to 
the Aether field is present. To maintain general covariance of Einstein 
gravity, we require that the preferred threading is dynamical. 
Since the timelike Aether field is nonvanishing at any spacetime points, 
it always breaks local Lorentz invariance.
In this sense, Einstein-Aether theory is distinguished from other 
Lorentz-violating theories restoring Lorentz invariance at some particular energy scales. 

The covariant action of Einstein-Aether theory, which was introduced by Jacobson and Mattingly  \cite{Jacobson:2000xp}, contains four 
derivative couplings of the Aether field with dimensionless coupling constants $c_{1,2,3,4}$ 
besides the Ricci scalar $R$. The unit vector constraint on the timelike 
Aether field can be incorporated into the action as a Lagrange multiplier 
of the form $\lambda (u^{\mu} u_{\mu}+1)$. 
We should mention that there was also an equivalent approach based on a tetrad formalism advocated by Gasperini \cite{Gasperini:1987nq}.
The Einstein-Aether framework can encompass several classes of vector-tensor theories such as the spontaneous breaking of Lorentz invariance in string theory \cite{Kostelecky:2008ts} and cuscuton theories 
with a quadratic scalar potential \cite{Afshordi:2006ad,Bhattacharyya:2016mah}. 
There are also extended versions of 
Einstein-Aether theory in which a symmetry-breaking potential for the vector is 
introduced \cite{Gripaios:2004ms} or the Aether coupling functions 
are generalized \cite{Zlosnik:2006zu,Kanno:2006ty}. 
The generalized Einstein-Aether theory of 
Ref.~\cite{Zlosnik:2006zu} is subject to severe constraints on the coupling functions, if 
$c_{1} + c_{3} \not = 0$ \cite{Chesler:2017khz}.

The perturbative analysis of Einstein-Aether theory on the Minkowski background (with all nonvanishing coupling constants $c_{1,2,3,4}$) shows that there are one scalar, two vector, and two tensor propagating degrees of freedom \cite{Jacobson:2004ts}. 
As we will review in Sec.~\ref{EAsec}, their squared propagation speeds
are given, respectively, by Eqs.~(\ref{cSm}), (\ref{cVm}), and (\ref{cTm}), 
all of which are different from that of light. 
The gravitational-wave event GW170817 of a black-hole (BH)-neutron star 
binary, along with the gamma-ray burst 170817A, put a stringent 
limit $|c_T-1| \lesssim 10^{-15}$ on the tensor propagation 
speed $c_T$ \cite{LIGOScientific:2017zic}, 
thereby translating to the bound 
$|c_1+c_3| \lesssim 10^{-15}$ \cite{Gong:2018cgj,Oost:2018tcv}. 
The coupling constants $c_{1,2,3,4}$ have been constrained from 
other experiments and observations such as gravitational 
Cerenkov radiation \cite{Elliott:2005va}, big-bang 
nucleosynthesis \cite{Carroll:2004ai}, solar-system tests 
of gravity \cite{Foster:2005dk}, binary pulsars \cite{Foster:2007gr,Yagi:2013qpa,Yagi:2013ava,Gupta:2021vdj}, 
and gravitational waveforms \cite{Hansen:2014ewa,Zhang:2019iim,Schumacher:2023cxh}.
Despite those numerous observational data, there are still wide regions 
of parameter space that are compatible with all these constraints. 

If we apply Einstein-Aether theory to the physics on 
a static and spherically symmetric (SSS) background, it is known that 
there are some nontrivial BH solutions endowed 
with the Aether hair \cite{Eling:2006ec,Foster:2005fr,Konoplya:2006rv,Garfinkle:2007bk,Berglund:2012bu,Gao:2013im,Lin:2014eaa,Ding:2015kba,Ding:2016wcf,Ding:2017gfw,Lin:2017cmn,Ding:2018whp,Chan:2019mdn,Zhu:2019ura,Chan:2020amr,Khodadi:2020gns,Zhang:2020too,Oost:2021tqi,Mazza:2023iwv}. 
Besides the usual metric horizon at which the time translation Killing 
vector $\zeta^{\mu}$ becomes null ($\zeta^{\mu}\zeta_{\mu}=0$), 
broken Lorentz invariance can give rise to the existence of 
a universal horizon at which the Aether field $u_{\mu}$ is orthogonal 
to $\zeta^{\mu}$, i.e., $u_{\mu}\zeta^{\mu}=0$ \cite{Blas:2011ni}. 
This universal horizon, which lies inside the metric horizon, can be
interpreted as a causal boundary of any speeds of propagation. 
In other words, once a wave signal is trapped inside the universal horizon, 
it does not escape from BHs toward spatial infinity. 
These distinguished features in Einstein-Aether theory may  
manifest themselves for inspiral gravitational waveforms emitted from 
BH binaries and BH quasinormal modes in the ringdown phase. 
Thus, the upcoming high-precision observations of gravitational waves 
will offer the possibility of probing the signature of 
BHs with the Aether hair.

The BH perturbation theory, which was originally developed by 
Regge-Wheeler \cite{Regge:1957td} and Zerilli \cite{Zerilli:1970se}, 
plays a crucial role in computing the quasinormal modes of BHs. 
Moreover, the linear stability of BHs is known by studying 
conditions for the absence of ghosts and Laplacian instabilities 
in the small-scale limit. In scalar-tensor Horndeski theories \cite{Horndeski:1974wa}, 
for example, the second-order actions of odd- and even-parity perturbations 
on the SSS background were derived 
in Refs.~\cite{Kobayashi:2012kh,Kobayashi:2014wsa,Kase:2021mix} 
for exploring the linear stability of hairy BHs.
In Refs.~\cite{Minamitsuji:2022mlv,Tsujikawa:2022lww,Kase:2023kvq}, 
it was found that the angular propagation speeds 
of even-parity perturbations, besides other stability conditions, 
are important to exclude a large class of hairy BHs due to 
Laplacian instabilities. 
As a result, the presence of a Gauss-Bonnet term coupled to 
the scalar field plays a prominent role in the realization of 
linearly stable BH solutions in Horndeski theories \cite{Minamitsuji:2022vbi}.

In Einstein-Aether theory, the second-order action of odd-parity 
perturbations was derived in Ref.~\cite{Tsujikawa:2021typ} by 
using a standard SSS coordinate 
introduced later in Eq.~(\ref{SSS}). 
The odd-parity sector contains two propagating degrees 
of freedom: (1) one tensor mode arising from the gravitational 
perturbation $\chi$, and (2) one vector mode arising from 
the Aether perturbation $\delta u$.
The no-ghost conditions and propagation speeds for 
$\chi$ and $\delta u$ were obtained by dealing with 
the $t$ coordinate as a time clock \cite{Tsujikawa:2021typ}.
In Einstein-Aether theory, however, there is a preferred timelike 
direction along the unit Aether field. 
Since the timelike property of $t$ coordinate is not always
ensured in this setup, the choice of $t$ and $r$ coordinates 
should not be necessarily appropriate for discussing 
the linear stability of BHs. 

On the SSS spacetime where the background 
Aether field does not have vorticity, it is possible to locally 
choose a timelike coordinate $\phi$ in the form 
$u_{\mu}|_{\rm background}=-\eta \partial_{\mu} \phi$, 
where $\eta$ is a nonvanishing function.
This scalar field $\phi$, which was named ``khronon'' 
in Ref.~\cite{Blas:2010hb}, defines the timelike direction in the foliation 
structure of spacetime. 
On the SSS background, 
one can introduce an Aether-orthogonal frame in which   
the Aether field is orthogonal to spacelike hypersurfaces. 
Indeed, it is known that \cite{Jacobson:2010mx} Einstein-Aether theory 
in such a configuration 
is equivalent to the infrared limit of the non-projectable version of 
Ho$\check{\rm r}$ava 
gravity \cite{Blas:2009qj}{\footnote{It should be noted that the equivalence between Einstein-Aether gravity and khronometric theory (or the infrared limit of non-projectable Ho$\check{\rm r}$ava gravity \cite{Horava:2009uw}) holds only when the Aether field has zero vorticity. In particular,  their Hamiltonian structures are different. In fact, while in  Einstein-Aether gravity there are five propagating local physical degrees of freedom, 
in khronometric theory the number of propagating 
local physical degrees of freedom is three \cite{Blas:2011ni,Lin:2017jvc}, so is 
in Ho$\check{\rm r}$ava gravity \cite{Wang:2017brl}.}.

Since the Aether-orthogonal frame is a proper choice of 
the timelike coordinate orthogonal to spacelike hypersurfaces, 
we will revisit the linear stability analysis of BHs in the odd-parity 
sector for this coordinate system. 
In Sec.~\ref{EAsec}, we briefly review current constraints on 
the coupling constants $c_{1,2,3,4}$ of derivative couplings 
of the Aether field. 
In Sec.~\ref{dissec}, we will see how a naive choice of 
the usual SSS coordinate (\ref{SSS})
can cause apparent instabilities and introduce the 
Aether-orthogonal frame as well as relations between 
two different frames. 
In Sec.~\ref{stasec}, we transform the second-order action of 
odd-parity perturbations derived in Ref.~\cite{Tsujikawa:2021typ} 
to that in the Aether-orthogonal frame and show that, for large radial and angular momentum modes, the no-ghost 
conditions and speeds of propagation are identical to those  
of vector and tensor perturbations on the Minkowski background. 
Thus, unlike the results in Ref.~\cite{Tsujikawa:2021typ}, 
the linear stability of BHs against odd-parity perturbations 
does not add new conditions to those known in the literature. 
Sec.~\ref{consec} is devoted to conclusions.

Throughout the paper, we will use the natural unit in which
the speed of light $c$ and the reduced Planck constant $\hbar$ 
are unity. We also adopt the metric signature $(-,+,+,+)$.

\section{Einstein-Aether theory and current constraints}
\label{EAsec}

Einstein-Aether theory is given by the action \cite{Jacobson:2000xp}
\be
{\cal S}=\frac{1}{16\pi G_{\ae}} \int  {\rm d}^4 x \sqrt{-g} 
\left[ R+{\cal L}_{\ae}+\lambda (g_{\mu \nu} u^{\mu} u^{\nu}+1) 
\right],
\label{action}
\ee
where $G_{\ae}$ is a constant corresponding to the gravitational coupling, 
$R$ is the Ricci scalar, $g$ is the determinant 
of metric tensor $g_{\mu \nu}$, 
$\lambda$ is a Lagrange multiplier, 
$u^{\mu}$ is the Aether vector field, and 
\be
{\cal L}_{\ae}=-{M^{\alpha \beta}}_{\mu \nu}
\nabla_{\alpha} u^{\mu} \nabla_{\beta} u^{\nu}\,,
\ee
with 
\be
{M^{\alpha \beta}}_{\mu \nu}:=c_1 g^{\alpha \beta} g_{\mu \nu}
+c_2 \delta^{\alpha}_{\mu} \delta^{\beta}_{\nu}
+c_3 \delta^{\alpha}_{\nu} \delta^{\beta}_{\mu}
-c_4 u^{\alpha} u^{\beta} g_{\mu \nu}\,.
\ee
The Greek indices run from 0 to 3, 
$\nabla_{\alpha}$ is a covariant derivative operator 
with respect to $g_{\mu \nu}$, and $c_{1,2,3,4}$  
are four dimensionless coupling constants. 

Varying the action (\ref{action}) with respect to $\lambda$, 
it follows that 
\be
\label{be1}
g_{\mu \nu} u^{\mu} u^{\nu}=-1\,.
\ee
This constraint ensures the existence of a timelike 
unit vector field at any spacetime points, 
so that there is a preferred threading responsible 
for the breaking of Lorentz invariance. 
Varying Eq.~(\ref{action}) with respect to $u^{\mu}$, we obtain
\be
\nabla_{\mu} {J^{\mu}}_{\alpha}+\lambda u_{\alpha}
+c_4 u^{\beta} \nabla_{\beta} u^{\mu} \nabla_{\alpha} u_{\mu}=0\,,
\label{be2}
\ee
where
\ba
{J^{\mu}}_{\alpha} &:=& 
{M^{\mu \nu}}_{\alpha \beta} 
\nabla_{\nu} u^{\beta}\,.
\label{Jmu}
\ea
Multiplying Eq.~(\ref{be2}) by $u^{\alpha}$ 
and using Eq.~(\ref{be1}), 
the Lagrange multiplier can be expressed as 
\be
\lambda=u^{\alpha} \nabla_{\mu} {J^{\mu}}_{\alpha}
+c_4 (u^{\beta} \nabla_{\beta} u^{\mu})
(u^{\rho} \nabla_{\rho} u_{\mu})\,. 
\label{Lagm}
\ee
The gravitational field equations derived 
by the variation of 
(\ref{action}) with respect to $g_{\mu \nu}$ are 
\ba
\hspace{-0.7cm}
G_{\alpha \beta}
&=&\nabla_{\mu} \left[ u_{(\alpha} {J^{\mu}}_{\beta )}
+u^{\mu}J_{(\alpha \beta)}-u_{(\alpha} {J_{\beta)}}^{\mu} \right]
\nonumber \\
\hspace{-0.7cm}
&&+ c_1 \left( \nabla_{\alpha} u^{\nu} \nabla_{\beta} u_{\nu} 
-\nabla^{\nu} u_{\alpha} \nabla_{\nu} u_{\beta} \right)
\nonumber \\
\hspace{-0.7cm}
&&
+c_4 (u^{\rho} \nabla_{\rho} u_{\alpha})
(u^{\nu} \nabla_{\nu} u_{\beta})\nonumber \\
\hspace{-0.7cm}
&&
+\frac{1}{2} g_{\alpha \beta}{\cal L}_{\ae}
+\lambda u_{\alpha} u_{\beta},
\label{Ein}
\ea
where $G_{\alpha \beta}$ is the Einstein tensor.

In general, the theory contains three different 
species of propagating degrees of freedoms, i.e., spin-0 (scalar), spin-1 (vector), and 
spin-2 (tensor) modes. 
On the Minkowski background with the line element 
\be
\rd s^2=\eta_{\mu \nu}\rd x^{\mu} \rd x^{\nu}=-\rd t^2+\delta_{ij}\rd x^i \rd x^j\,,
\label{Min}
\ee
the Aether field is aligned along the $t$ direction, as $u^{\mu}=\delta^{\mu}_0$.
According to the perturbative analysis on the background (\ref{Min}), 
the squared propagation speeds of spin-0, 
spin-1, and spin-2 modes 
are given, respectively, by \cite{Jacobson:2004ts,Oost:2018tcv}
\ba
c_S^2 & = & \frac{c_{123}(2-c_{14})}{c_{14}(1-c_{13}) (2+c_{13} + 3c_2)}\,,\label{cSm}\\
c_V^2 & = & \frac{2c_1 -c_{13} (2c_1-c_{13})}{2c_{14}(1-c_{13})}\,,\label{cVm}\\
c_T^2 & = & \frac{1}{1-c_{13}}\,,\label{cTm}
\ea
where $c_{ij}:=c_i+c_j$ and $c_{ijk}:=c_i + c_j + c_k$. 
The coefficients of the kinetic terms for each mode are 
\ba
q_S & = & \frac{(1-c_{13})(2+c_{13}+3c_2)}{c_{123}} \,,
\label{qS}\\
q_V & = & c_{14} \,,\\
q_T & = & 1-c_{13} \,.
\label{qT}
\ea
So long as the denominators in Eqs.~(\ref{cSm})-(\ref{qS}) 
do not vanish, there are one scalar, two vector, and two tensor propagating degrees of freedom in general.

If we require that the theory:  
(i)  be self-consistent, such as free of ghosts 
and Laplacian instabilities; and 
(ii) be compatible with all the experimental 
and observational constraints obtained so far, it was found that the coupling 
constants must satisfy the following conditions \cite{Oost:2018tcv}
\ba
\label{CD1a}
&& \left|c_{13}\right| \lesssim 10^{-15}\,,\\
\label{CD1b}
&& 0 < c_{14} \leq 2.5 \times 10^{-5}\,,\\
\label{CD1c}
&&  c_{14} \leq c_2 \leq  0.095\,,\\
\label{CD1d}
&&  c_{4} \leq 0\,.
\ea
It should be noted that the recent studies of the neutron star binary systems showed that one 
of the parameterized post-Newtonian parameters, $\alpha_1=-4c_{14}$, is further restricted to 
$|\alpha_1| < 10^{-5}$ \cite{Gupta:2021vdj}. 
This translates to the limit
\be
\label{CD1ba}
0 \lesssim c_{14} \lesssim  2.5 \times 10^{-6}, 
\ee
which is stronger than the bound derived from lunar laser ranging experiments by one order of magnitude \cite{Oost:2018tcv}. 

%%%%%%%%%%%%%%%%%%%%%%%%%%%%%%%%%%%
\section{Disformal transformation and Aether-orthogonal frame}
\label{dissec}
%%%%%%%%%%%%%%%%%%%%%%%%%%%%%%%%%%%

%
\subsection{Disformal transformation}

Under a redefinition of the metric accompanied 
with the Aether field in the form 
$\tilde{g}_{\mu \nu}=g_{\mu \nu}+B u_{\mu}u_{\nu}$, where $B$ is a constant, the structure of the action (\ref{action}) is preserved 
with a change of the coupling constants 
$\tilde{c}_{1,2,3,4}$ in the transformed 
frame \cite{Foster:2005ec}. 
This redefinition stretches the metric tensor in the Aether direction by a factor $1-B$. On choosing $B=1-c_I^2$ for the Minkowski metric $g_{\mu \nu}=\eta_{\mu \nu}$, 
where the subscript $I$ is either $S, V, T$ with the squared propagation speeds $c_I^2$ given by 
Eqs.~(\ref{cSm})-(\ref{cTm}), it is possible to transform to a metric frame 
$\tilde{g}_{\mu \nu}$ in which 
one of the speeds is equivalent to 1 \cite{Eling:2006ec,Jacobson:2007veq}.

One can perform a more general disformal transformation \cite{Bekenstein:1992pj} of the form
\be
\bar{g}_{\mu\nu} = \Omega^2 \left(g_{\mu\nu} + B u_{\mu}u_{\nu}\right)\,,
\label{eqn:disformal-tr-metric}
\ee
where the conformal factor $\Omega$ and 
the disformal factor $B$ are constants, and 
the Aether field $u_{\mu}$ satisfies 
the unit-vector constraint (\ref{be1}).
The corresponding inverse metric 
and determinant are
\ba
 \bar{g}^{\mu\nu} &=& \frac{1}{\Omega^2}\left(g^{\mu\nu} -\frac{B}{1-B}u^{\mu}u^{\nu}\right),\nn\\
 \sqrt{-\bar{g}} &=& \Omega^4\sqrt{-(1-B)g}\,,
\ea
where we are assuming that $1-B>0$. 
The Aether field has a different (but constant) norm with respect to $\bar{g}^{\mu\nu}$ as
\be
\bar{g}^{\mu\nu}u_{\mu}u_{\nu} 
= -\frac{1}{\Omega^2 (1-B)}\,.
\ee
Hence, it makes sense to define
\be
\bar{u}_{\mu} = \Omega\sqrt{1-B}\,u_{\mu}\,, \qquad 
\bar{u}^{\mu} = \bar{g}^{\mu\nu}\bar{u}_{\nu}\,.
\ee

The first covariant derivative of the Aether field with respect to $\bar{g}_{\mu\nu}$ is given by \cite{Domenech:2018vqj} 
\be
\bar{\nabla}_{\mu}u_{\nu} = \nabla_{\mu}u_{\nu} 
- B \delta\Gamma^{\rho}_{\mu\nu}u_{\rho}\,,
\end{equation}
where
\ba
\delta\Gamma^{\rho}_{\mu\nu} &=& 
u_{(\mu}F_{\nu)}^{\ \ \rho} 
+ \frac{u^{\rho}}{1-B} 
\left[ \nabla_{(\mu}u_{\nu)} 
+Bu_{\lambda}u_{(\mu}\nabla^{\lambda}u_{\nu)}
\right]\,,\nn\\
\quad F_{\mu\nu} &=& \nabla_{\mu}u_{\nu} - \nabla_{\nu}u_{\mu}\,.
\ea
The Einstein-Hilbert action transforms 
as~\cite{Domenech:2018vqj}
\begin{widetext}
\begin{align}
 \int \rd^4x\sqrt{-\bar{g}}\,\bar{R} 
 =& \int \rd^4x\sqrt{-g} 
 \left[ \Omega^2 \sqrt{1-B} R 
 - \frac{\Omega^2B}{\sqrt{1-B}}
 \left\{ (\nabla_{\mu}u^{\mu})^2 
- \nabla_{\rho}u^{\sigma}\nabla_{\sigma}u^{\rho}\right\} \right.\nonumber\\
& \left. 
+ \frac{\Omega^2B^2}{2\sqrt{1-B}}\left(u^{\mu}u^{\nu}
F_{\mu\rho}F_{\nu}^{\ \rho} + \frac{1}{2}F_{\mu\nu}F^{\mu\nu}\right)\right]\,.
\end{align}
\end{widetext}
On using these relations, in the absence of matter fields, Einstein-Aether theory for the combination 
($g_{\mu\nu}$, $u_{\mu}$) with the constant parameters ($G_{\ae}$, $c_1$, $c_2$, $c_3$, $c_4$) is equivalent to Einstein-Aether theory for the combination 
($\bar{g}_{\mu\nu}$, $\bar{u}_{\mu}$) with a different set of parameters 
($\bar{G}_{\ae}$, $\bar{c}_1$, $\bar{c}_2$, $\bar{c}_3$, $\bar{c}_4$), 
where \cite{Foster:2005ec}
\ba
& &
 \bar{G}_{\ae} = \Omega^2 \sqrt{1-B}\,{G}_{\ae}\,, 
 \quad 
 \bar{c}_{14} = c_{14},\nn\\
& &
 \bar{c}_{123}= (1-B)c_{123}\,, \quad 
 \bar{c}_{13}-1 = (1-B)(c_{13} -1)\,, \nn\\
& &
 \bar{c}_{1}- \bar{c}_{3}-1= \frac{1}{1-B}\left(c_1-c_3-1 \right) \,. 
 \label{eqn:disformal-tr-parameters}
\ea
More explicitly, the coefficients ${\bar c}_i$ are related to $c_i$, as 
\ba
\bar{c}_1 &=& \frac{2 c_1-2 (c_1+c_3)B+(c_1+c_3-1)B^2}{2(1-B)}\,, 
\label{barc1}
\\
\bar{c}_2 &=& c_2 (1-B)-B\,,
 \\
\bar{c}_3 &=& \frac{2 c_3-(c_1+c_3-1)B(2-B)}{2(1-B)}\,, 
 \\
\bar{c}_4 &=& \frac{2c_4+2(c_3-c_4)B-(c_1+c_3-1)B^2}{2(1-B)}\,.
\label{barc4}
\ea
We consider the Minkowski background characterized by the metric 
tensor $g_{\mu \nu}=\eta_{\mu \nu}$ and perform 
the disformal transformation 
(\ref{eqn:disformal-tr-metric}). 
Upon choosing
\be
B=1-c_I^2\,,\quad {\rm where} \quad 
I=S,V,T\,,
\ee
and using Eqs.~(\ref{cSm})-(\ref{qT}) and 
Eqs.~(\ref{barc1})-(\ref{barc4}), 
the squared propagation speeds $\bar{c}_I^2$ in the frame ($\bar{g}_{\mu\nu}$, $\bar{u}_{\mu}$) yield
\be
\bar{c}_I^2=1\,, 
\label{cI}
\ee
for each subscript $I=S,V,T$. Furthermore, the coefficients of the time kinetic terms yield
\be
\bar{Q}_I = c_I^2Q_I\,,
\ee
where\footnote{For the vector perturbation, the no-ghost condition changes from $q_V>0$ to $Q_V>0$ if one swaps the roles of the dynamical variable and its canonical momentum by a canonical transformation. See e.g., Appendix B of \cite{DeFelice:2015hla} or/and Section IV of \cite{Gumrukcuoglu:2016jbh} for a technique to perform canonical transformations at the level of the Lagrangian.}
\be
Q_S \equiv q_S\,, \qquad 
Q_V \equiv \frac{q_V}{c_V^2}\,, \qquad 
Q_T \equiv q_T\,. 
\ee
Here, we have assumed 
\be
c_I^2>0\,,
\label{cI}
\ee
in order to avoid the gradient instability of perturbations. Under this assumption, 
the inequality $1-B>0$ holds and thus the 
disformal transformation does not change the Lorentzian signature of the metric. 

\subsection{Aether-orthogonal frame}

If the background Aether field has zero vorticity, which is the case for 
any spherically symmetric configurations, 
one can locally choose the time coordinate $\phi$ such that 
\be
 \left. u_{\mu}\right|_{\rm background} = 
 -\eta \partial_{\mu} \phi = -\eta  \delta^{\phi}_{\mu}\,, 
 \label{muAe}
\ee
where $\eta$ is a nonvanishing function.
This choice of the time coordinate $\phi$ for the 
metric frame $g_{\mu \nu}$ is called 
the Aether-orthogonal frame. 
The unit-vector constraint (\ref{be1}) gives 
$\left. g^{\phi\phi}\right|_{\rm background} = -\eta^{-2}$ and
$\left. u^{\phi}\right|_{\rm background}=\eta^{-1}$.

At each point of physical interest, one can choose a local Lorentz frame and then perform a local Lorentz transformation so that the metric and Aether field are of the form 
\be \label{eqn:localbackground}
g_{\mu\nu}|_{\rm local}
 =-\eta^2 {\rm d}\phi^2 
 + \delta_{ij} {\rm d}x^i {\rm d}x^j\,, 
\quad \quad u_{\mu}|_{\rm local}=-\eta \delta_{\mu}^{\phi}\,.
\ee
At leading order in the geometrical optics approximation, i.e., for modes whose wavelengths are much shorter than the time and length scales of the background, the background in the vicinity of the point of interest can be approximated 
by Eq.~\eqref{eqn:localbackground}. 
In particular, in the vicinity of the point of interest, one can decompose the perturbations into spin-$0$ (scalar, $I=S$), spin-$1$ (vector, $I=V$) and spin-$2$ (tensor, $I=T$) modes. 
Let us consider the perturbations $\delta \chi_I$ corresponding to the $I$-excitation ($I=S,V,T$). Then, by definition of $Q_I$ and $c_I^2$, the kinetic and gradient terms of $\delta \chi_I$ should locally have the following structure
\be \label{eqn:localLkin}
L_{{\rm kin}, I} = \frac{1}{2}C_I Q_I \left[(\eta^{-1}\partial_{\phi} \delta\chi_I)^2 - c_I^2 \delta^{ij}\partial_i \delta \chi_I
\partial_j \delta\chi_I \right] + \cdots\,,
\ee
where $C_I$ are positive definite coefficients, which should not be confused with $C_1,C_2,\cdots$ introduced later in Sec.~\ref{stasec}, and $\cdots$ 
represents higher-order terms in the geometrical optics approximation. By undoing the local Lorentz transformation and going back to the original coordinate system before choosing the local Lorentz frame, the leading kinetic and gradient terms \eqref{eqn:localLkin} can be written in a general coordinate system as
\be \label{eqn:generalLkin}
L_{{\rm kin}, I} = -\frac{1}{2} C_I\bar{Q}_I \bar{g}_I^{\mu\nu}
\partial_{\mu} \delta \chi_I \partial_{\nu}  \delta \chi_I + \cdots\,,
\ee
at leading order in the geometrical optics approximation, where $\cdots$ again represents 
higher-order terms in the geometrical optics approximation. We have assumed that the time 
and spatial scales involved in the coordinate transformation between the original coordinate system and the local Lorentz frame in the vicinity of the point of interest are sufficiently longer than the wavelengths of the modes of interest. 

The local structure \eqref{eqn:generalLkin} clearly states that the perturbations $\delta \chi_I$ in the Aether-orthogonal frame do not behave as ghosts so long as $Q_I>0$. Indeed, if we choose the time coordinate $\phi$ in the Aether-orthogonal frame, then the leading time kinetic term is 
\ba
L_{{\rm kin}, I} &\ni& -\frac{1}{2}C_I\bar{Q}_I\bar{g}_I^{\phi \phi}(\partial_{\phi} \delta \chi_I)^2 +\cdots\nonumber\\
&=& \frac{C_I c_I^2 Q_I}{2\Omega_I^2\eta^2}(\partial_{\phi} \delta \chi_I)^2 + \cdots\,,
\label{Lkin}
\ea
which is positive. 

Rigorously speaking, we have only given a heuristic argument for the local structure \eqref{eqn:generalLkin} without a proof, which requires analysis similar to that 
in Ref.~\cite{Kubota:2022lbn}. In the rest of the present paper, we shall show that the kinetic and gradient terms for odd-parity perturbations (including $I=V$ and $I=T$ modes) around spherically symmetric BHs indeed have the local structure \eqref{eqn:generalLkin}. The same analysis for even-parity perturbations (including $I=S,V,T$ modes) will be left for a future work. 

\subsection{Spherically symmetric background and 
apparent instabilities}
\label{inssec}

Let us consider the SSS
background given by the line element 
\be
\rd s^2 = g_{\mu\nu} \rd x^{\mu} \rd x^{\nu} 
= -f(r) \rd t^2 + \frac{\rd r^2}{h(r)} + r^2 \Omega_{pq}
\rd \vartheta^p \rd\vartheta^q\,, \label{SSS}
\ee
together with the Aether-field configuration
\be
u^{\mu}\partial_{\mu} = 
a(r)\partial_t + b(r)\partial_r\,,
\label{Aecon}
\ee
where $f,h,a,b$ are functions of $r$. 
In Eq.~(\ref{SSS}) the angular part contains two 
angles $\theta$ and $\varphi$, such that 
$\Omega_{pq}\rd \vartheta^p \rd\vartheta^q
=\rd \theta^2+\sin^2 \theta\,\rd\varphi^2$.

The unit-vector constraint (\ref{be1}) gives the following relation 
\be
b=\epsilon \sqrt{(a^2 f -1)h}\,,
\label{bso}
\ee
where $\epsilon= \pm 1$. 
The existence of the Aether-field profile (\ref{bso}) with $b \neq 0$ requires 
that $(a^2 f -1)h>0$.

A metric (Killing) horizon is defined at the radius 
$r=r_{\rm g}$ at which the time translation 
Killing vector $\zeta^{\mu}=\delta^{\mu}_t$ 
is null, i.e., $\zeta^{\mu}\zeta_{\mu}=0$. 
This translates to the condition
\be
f(r_{\rm g})=0\,.
\ee
So long as $h$ also vanishes on the metric horizon,  
the product $a^2 f h$ is at least 
a positive constant at $r=r_{\rm g}$ to 
satisfy the condition $(a^2 f -1)h>0$.
This means that, as $r \to r_{\rm g}^{+}$, 
the temporal Aether-field component diverges as 
$a^2 \propto (fh)^{-1}$ 
for the coordinate system (\ref{SSS}).

The leading-order kinetic and gradient terms of 
perturbations $\delta \chi_I$ (where $I=S,V,T$) for the effective metric 
$\bar{g}^I_{\mu \nu}=\Omega_I^2 
[g_{\mu \nu} + (1-c_I^2) u_{\mu} u_{\nu}]$ 
(under the geometric optics approximation) are
\begin{widetext}
\begin{align}
L_{{\rm kin}, I} & = -\frac{1}{2}C_I\bar{Q}_I\bar{g}_I^{\mu\nu}
\partial_{\mu} \delta \chi_I \partial_{\nu}  \delta \chi_I= 
\frac{C_I\bar{Q}_I}{2\Omega_I^2}\left[ \frac{1}{f}(\partial_t \delta \chi_I )^2 - h(\partial_r \delta \chi_I )^2
 - \frac{1}{r^2}\Omega^{pq}\partial_p \delta \chi_I \partial_q \delta \chi_I 
 + \frac{1-c_I^2}{c_I^2}\left(a\partial_t \delta \chi_I +b\partial_r \delta \chi_I \right)^2\right]\nonumber\\
& \ni \frac{1}{2} f C_I Q_I^{\rm apparent}\left[\frac{1}{f}(\partial_t \delta \chi_I )^2 
- (c_{I,\Omega}^{\rm apparent})^2
\frac{1}{r^2}\Omega^{pq}\partial_p \delta \chi_I \partial_q \delta \chi_I \right]\,,
\end{align}
\end{widetext}
where
\ba
Q_I^{\rm apparent} &=& \frac{\bar{Q}_I}{\Omega_I^2}\left(\frac{1}{f} + \frac{1-c_I^2}{c_I^2}a^2\right)\,,
\label{qI} \\
\left(c_{I,\Omega}^{\rm apparent}\right)^2 &=& \left( 1 + \frac{1-c_I^2}{c_I^2}a^2f\right)^{-1}\,.
\label{cI}
\ea
Therefore, the coefficient $Q_I^{\rm apparent}$ of the kinetic term with respect to the Killing time $t$ 
may become negative, despite the fact that the kinetic term with respect to the time coordinate 
in the Aether-orthogonal frame is always positive as in Eq.~(\ref{Lkin}).
Also, the apparent angular sound speed squared $(c_{I,\Omega}^{\rm apparent})^2$ 
would have nontrivial position dependence, although the propagation speed squared $c_I^2$ 
relative to the Aether-orthogonal frame is a constant given by the theory. 

These behaviors are due to the deviation of the Killing time slicing from the Aether-orthogonal frame. The sound cones are not only narrowed or widened but also tilted relative to the Killing time slicing\footnote{See Ref.~\cite{Mukohyama:2002vq} for similar behaviors of sound cones for open string modes in the context of inhomogeneous tachyon condensation in string theory.}. On the other hand, the sound cones are not tilted relative to the Aether-orthogonal frame. 

Near the metric horizon $r=r_{\rm g}$, we have the following expansions
\begin{equation}
f = f_1(r-r_{\rm g}) + \cdots\,, \quad 
a = -\frac{1}{2f_1 \alpha_0}
(r-r_{\rm g})^{-1} + \cdots\,,
\label{fa}
\end{equation}
where $f_1$ and $\alpha_0$ are constants \cite{Tsujikawa:2021typ}. 
Note that $\alpha_0$ is the value of $\alpha$ 
at $r=r_{\rm g}$, where $\alpha$ is defined 
later in Eq.~(\ref{ab}).
In this regime, the quantities (\ref{qI}) and (\ref{cI}) can be estimated as\footnote{For $I=V$ and $c_{13}=0$, the result is to be compared with (5.32) and (5.33) of Ref.~\cite{Tsujikawa:2021typ} up to an arbitrary positive overall factor for $q_V^{\rm apparent}$.}
\begin{align}
Q_I^{\rm apparent} &= \frac{\bar{Q}_I}
{4f_1^2 \alpha_0^2\Omega_I^2}\frac{1-c_I^2}{c_I^2}(r-r_{\rm g})^{-2} + \cdots\,, \\
(c_{I,\Omega}^{\rm apparent})^2 &= 
4f_1 \alpha_0^2 \frac{c_I^2}{1-c_I^2}(r-r_{\rm g}) + \cdots\,.
\end{align}
Therefore, if $c_I^2>1$, the apparent no-ghost condition is violated near the horizon. 
Moreover, the apparent angular sound speed squared vanishes at the horizon. 
Let us stress again that the kinetic term with respect to the time coordinate in the Aether-orthogonal frame 
is always positive for $Q_I>0$ and that the propagation speed squared $c_I^2$ relative to 
the Aether-orthogonal frame is constant given by the theory. 
In Sec.~\ref{stasec}, we will address the linear stability of BHs in Einstein-Aether theory 
by choosing the Aether-orthogonal frame as
a timelike coordinate.

\subsection{Aether-orthogonal frame on the spherically symmetric background}
\label{Aeorsec}

The SSS background is described by the line element (\ref{SSS}) with the Aether-field profile (\ref{Aecon}). Alternatively, we can choose the following Eddington-Finkelstein 
coordinate \cite{Zhang:2020too}
\be
{\rm d}s^{2} =-f(r) {\rm d}v^{2} +2B(r) \rd v \rd r+ r^2 \Omega_{pq}
\rd \vartheta^p \rd\vartheta^q\,,
\label{metric2}
\ee
which is related to the coordinate (\ref{SSS}) as 
\be
\rd v=\rd t+\frac{\rd r}{\sqrt{fh}}\,,\qquad 
B(r)=\sqrt{\frac{f}{h}}\,.
\label{rdv}
\ee
The transformation to the coordinate (\ref{metric2}) is 
valid in the regime $f/h>0$, i.e., for the same 
signs of $f$ and $h$.
On this background, we take the Aether-field configuration 
\be
u^{\mu}\partial_{\mu} =-\alpha(r) \partial_v 
-\beta(r) \partial_r\,.
\label{Aecon2}
\ee
The $r$-dependent functions $\alpha$ 
and $\beta$ are related to $a$ and $b$ in Eq.~(\ref{Aecon}), as
\be
a+\frac{b}{\sqrt{fh}}=-\alpha\,,\qquad 
b=-\beta\,.
\label{ab}
\ee
From Eq.~(\ref{bso}), we obtain the following relations 
\be
a=-\frac{1+f\alpha^2}{2f \alpha}\,,\qquad 
b=\sqrt{fh} \frac{1-f\alpha^2}{2f\alpha}\,.
\label{ab2}
\ee
The sign of $b$ depends on that of 
$(1-f \alpha^2)/(f\alpha)$.
Then, the Aether field $u^{\mu}$ has nonvanishing components
\be
u^{v}=-\alpha\,,\qquad u^{r}= \sqrt{fh} \frac{1-f\alpha^2}{2f\alpha}\,.
\ee
Even though $a$ is divergent on the metric horizon 
($r=r_{\rm g}$), the quantity $\alpha$ can take 
a finite value $\alpha_0$ due to the relation $fa=-1/(2\alpha_0)$ at 
$r=r_{\rm g}$, see Eq.~(\ref{fa}).
On using the metric components $g_{vv}=-f$, $g_{vr}=g_{rv}=B$, and $g_{rr}=0$, 
the nonvanishing components of $u_{\mu}$ are 
$u_v=f\alpha-B\beta$ and $u_r=-B \alpha$, so that 
\be
u_{v}=\frac{1+f\alpha^2}{2\alpha}\,,\qquad 
u_{r}= -\alpha \sqrt{\frac{f}{h}}\,.
\ee
A vector field $s_{\mu}$ that is orthogonal to $u^{\mu}$ obeys 
the relation $s_{\mu} u^{\mu}=0$. 
This has the following nonvanishing components
\be
s_v=\frac{1-f\alpha^2}{2\alpha}\,,\qquad 
s_r=\alpha \sqrt{\frac{f}{h}}\,.
\ee
Moreover, it satisfies the relation $s_{\mu}s^{\mu}=1$. 

Now, we introduce the two coordinates $\phi$ 
and $\psi$, as
\ba
\rd \phi &=& \rd v+\frac{u_r}{u_v} \rd r
=\rd v-\frac{2\alpha^2}{1+f \alpha^2} \sqrt{\frac{f}{h}}\,\rd r\,,
\label{rdphi}\\
\rd \psi &=& -\rd v-\frac{s_r}{s_v} \rd r
=-\rd v-\frac{2\alpha^2}{1-f \alpha^2} \sqrt{\frac{f}{h}}\,\rd r,
\label{rdpsi}
\ea
which mean that $\phi$ is constant on a hypersurface 
orthogonal to $u_{\mu}$ and that $\psi$ is constant on a hypersurface orthogonal to $s_{\mu}$. 
Note that $s_v$ goes to 0 as $r \to \infty$ and 
hence Eq.~(\ref{rdpsi}) is valid for a finite 
distance $r$.  
Since we already know the linear stability conditions in the asymptotically flat 
regime \cite{Jacobson:2004ts,Oost:2018tcv},
it is sufficient to focus on the stability 
in the region with finite $r$. 
If we introduce the coordinate $\tilde{\psi}$ 
as $\rd\tilde{\psi}=-s_v \rd v-s_r  \rd r$ 
instead of $\psi$ to avoid the divergence 
of $s_r/s_v$ at spatial infinity, then $\tilde{\psi}$ is not integrable due to the $r$ 
dependence in $s_v$. In this sense 
we choose the coordinate system 
$(\phi, \psi)$, which satisfies the  
integrability condition.

Since $\partial_{\mu} \phi=\delta^{v}_{\mu}-2\alpha^2/(1+f\alpha^2)\sqrt{f/h}\, 
\delta^{r}_u$ from Eq.~(\ref{rdphi}), the background Aether field 
can be expressed in the form 
\be
u_{\mu}=\frac{1+f \alpha^2}{2\alpha} \partial_{\mu} \phi\,,
\ee
and hence $\eta=-(1+f \alpha^2)/(2\alpha)=fa$ in Eq.~(\ref{muAe}). 
Thus, the coordinate system $(\phi, \psi)$ corresponds to 
the Aether-orthogonal frame in which 
$\phi$ is the time measured by observers comoving 
with the Aether field. 

Substituting the first of Eq.~(\ref{rdv}) into Eqs.~(\ref{rdphi}) 
and (\ref{rdpsi}), the relation between the Aether-orthogonal frame 
and the coordinate (\ref{SSS}) is given by \cite{Lin:2014eaa}
\ba
\label{eq2.22}
\rd \phi &=& \rd t +\frac{1}{\sqrt{fh}} \frac{1-f \alpha^2}{1+f \alpha^2} \rd r\,,
\label{rdphi2} \\
\rd \psi &=& - \rd t -\frac{1}{\sqrt{fh}} \frac{1+f \alpha^2}{1-f \alpha^2} \rd r\,.
\label{rdpsi2}
\ea
Solving these equations for $\rd t$ and $\rd r$ and substituting them into 
Eq.~(\ref{SSS}), the line element is expressed as 
\be
\rd s^{2} =-\frac{\left( 1+f\alpha^2 \right)^2}{4\alpha^2} \rd\phi^2 
+ \frac{\left(1-f \alpha^2 \right)^2}{4\alpha^2} \rd\psi^2
+r^2 \Omega_{pq} \rd \vartheta^p \rd\vartheta^q.
\label{phipsi}
\ee
Since $g_{\phi\phi}$ is always non-positive, 
$\phi$ is a timelike coordinate. 
Similarly, $\psi$ is a spacelike coordinate 
due to the positivity of $g_{\psi\psi}$.
Note that a universal horizon is defined as  
the radius $r_{\rm UH}$ at which the time translation Killing vector $\zeta^{\mu}$ is orthogonal to 
$u_{\mu}$, i.e., $\zeta^{\mu}u_{\mu}=0$. 
This corresponds to the radius satisfying
\be
\left( 1+f\alpha^2 \right)|_{r = r_{\text{UH}}} = 0\,,
\label{uni}
\ee
at which the metric component $g_{\phi \phi}$ in 
Eq.~(\ref{phipsi}) is vanishing.
It is clear that Eq.~(\ref{uni}) has solutions 
only when $f < 0$, that is, the universal horizon always exists inside the metric horizon. 
For more details, see Ref.~\cite{Lin:2014eaa}.

{}From Eqs.~(\ref{rdphi2}) and (\ref{rdpsi2}), we have 
\ba
& &
\frac{\partial \phi}{\partial t}=1\,,\qquad 
\frac{\partial \phi}{\partial r}=\frac{1}{\sqrt{fh}} \frac{1-f \alpha^2}{1+f \alpha^2}\,,
\\
& &
\frac{\partial \psi}{\partial t}=-1\,,\qquad 
\frac{\partial \psi}{\partial r}=-\frac{1}{\sqrt{fh}} \frac{1+f \alpha^2}{1-f \alpha^2}\,.
\ea
Then, for a given function ${\cal F}$ of $t$ and $r$, the $t$ and $r$ 
derivatives of ${\cal F}$ are given, respectively, by 
\ba
\hspace{-0.5cm}
& &
\frac{\partial {\cal F}}{\partial t}
={\cal F}_{,\phi}-{\cal F}_{,\psi}\,,
\label{calF1}\\
\hspace{-0.5cm}
& &
\frac{\partial {\cal F}}{\partial r}=\frac{1}{\sqrt{fh}} \left( 
\frac{1-f\alpha^2}{1+f \alpha^2}{\cal F}_{,\phi}
-\frac{1+f\alpha^2}{1-f \alpha^2}{\cal F}_{,\psi} 
\right)\,,\label{calF2}
\ea
where ${\cal F}_{,\phi}:=
\partial {\cal F}/\partial \phi$ and 
${\cal F}_{,\psi}:=\partial {\cal F}/\partial \psi$. 
The relations (\ref{calF1}) and (\ref{calF2})
will be used in Sec.~\ref{stasec}.

%%%%%%%%%%%%%%%%%%%%%%%%%%%%%%%%%%%
\section{Odd-parity stability in the 
Aether-orthogonal frame} 
\label{stasec}
%%%%%%%%%%%%%%%%%%%%%%%%%%%%%%%%%%%

In this section, we study the linear stability 
of SSS BHs against odd-parity perturbations 
in the Aether-orthogonal frame. The second-order action in the odd-parity sector   
was derived in Ref.~\cite{Tsujikawa:2021typ} 
for the line element (\ref{SSS}). 
Since the constant $t$ hypersurfaces are not always spacelike, 
the coordinate choice (\ref{SSS}) is 
not suitable for studying the linear stability 
of BHs. 
We will express the second-order 
action of odd-parity perturbations by using 
the derivatives with respect to $\phi$ and $\psi$. 
In the following, we will discuss the two cases: 
(A) $l \geq 2$, and (B) $l=1$, in turn, where 
$l$'s are spherical multipoles.

\subsection{$l \geq 2$}

On the SSS background (\ref{SSS}) with 
$\Omega_{pq}\rd \vartheta^p \rd\vartheta^q
=\rd \theta^2+\sin^2 \theta\,\rd\varphi^2$, 
metric perturbations $h_{\mu \nu}$ can be separated 
into odd-parity (axial) and even-parity 
(polar) sectors \cite{Regge:1957td,Zerilli:1970se}. 
We express $h_{\mu \nu}$ in terms of the 
spherical harmonics $Y_{lm}(\theta, \varphi)$. 
We will focus on axial perturbations with 
the parity $(-1)^{l+1}$.
We choose the gauge in which the components $h_{ab}$ 
vanish, where the subscripts $a$ and $b$ 
denote either $\theta$ or $\varphi$.
Then, the nonvanishing components of odd-parity 
metric perturbations are given by 
\ba
h_{ta} &=& \sum_{l,m} Q_{lm} (t,r) E_{ab} \nabla^b Y_{lm} 
(\theta, \varphi)\,,\\
h_{ra} &=& \sum_{l,m} W_{lm} (t,r) E_{ab} \nabla^b Y_{lm} 
(\theta, \varphi)\,,
\ea
where $Q_{lm}$ and $W_{lm}$ are functions of $t$ and $r$.
The tensor $E_{ab}$ is antisymmetric with nonvanishing components $E_{\theta \varphi}=-E_{\varphi \theta}
=\sin \theta$.

In the odd-parity sector, the Aether field has the following components
\ba
& &
u_t=-a(r) f(r)\,,\qquad 
u_r=\frac{b(r)}{h(r)}\,, \nn\\
& &
u_a = \sum_{l,m} \delta u_{lm} (t,r) E_{ab} 
\nabla^b Y_{lm} (\theta, \varphi)\,, 
\label{umuA}
\ea
where $\delta u_{lm}$ is a function of $t$ and $r$. 

In the following, we will set $m=0$ without loss of generality. We also omit the subscripts $l$ and $m$ 
from the perturbations $Q_{lm}$, $W_{lm}$, 
and $\delta u_{lm}$.
Expanding the action (\ref{action}) 
up to quadratic order and integrating it with respect to $\theta$ and $\varphi$, we obtain the second-order action 
containing the fields $Q$, $W$, $\delta u$ 
and their $t, r$ derivatives. The dynamical field associated with 
the gravitational (tensor) perturbation 
is given by \cite{Tsujikawa:2021typ}
\be
\chi:= \dot{W}-Q'+\frac{2}{r}Q
+\frac{C_2 \dot{\delta u}+C_3 \delta u'+C_4 \delta u}{C_1}\,,
\label{chidef}
\ee
where a dot and prime represent the derivatives 
with respect to $t$ and $r$, respectively, and
\ba
\hspace{-0.7cm}
& &
C_1=\frac{(1-c_{13})h}{2r^2 f}\,,\quad
C_2=-\frac{c_{13}b}{2r^2 f}\,,\quad
C_3=-\frac{c_{13}ah}{2r^2},\nonumber \\
\hspace{-0.7cm}
& &
C_4=\frac{[(2c_{14}-c_{13})(fa'+af')r+2c_{13}af]h}
{2r^3 f}\,.
\ea
Taking into account the field $\chi$ as a form 
of the Lagrange multiplier and varying 
the corresponding second-order action 
with respect to $W$ and $Q$, 
we can eliminate $W$, $Q$, and their derivatives 
from the action. 
Then, the resulting quadratic-order action can be 
expressed in the form \cite{Tsujikawa:2021typ}
\be
{\cal S}_{\rm odd}=\sum_{l}L \int 
{\rm d}t {\rm d} r\,{\cal L}_{\rm odd}\,,
\label{Sodd}
\ee
where 
\be
L=l(l+1)\,,
\ee
and 
\ba
{\cal L}_{\rm odd}
&=&\frac{r^2}{16 \pi G_{\ae}} \sqrt{\frac{f}{h}} \;\;
(\dot{\vec{\mathcal{X}}}^{t}{\bm K}\dot{\vec{\mathcal{X}}}
+\dot{\vec{\mathcal{X}}}^{t}{\bm R}\vec{\mathcal{X}}'
\nonumber \\
&&\qquad \qquad \qquad 
+\vec{\mathcal{X}}'^{t}{\bm G}\vec{\mathcal{X}}' 
+\vec{\mathcal{X}}^{t}{\bm M}\vec{\mathcal{X}})\,.
\label{Lodd}
\ea
The vector field $\vec{\mathcal{X}}$ is given by 
\be
\vec{\mathcal{X}}=
\left(\begin{array}{c} \chi\\
\delta u
\end{array}
\right), \quad 
\vec{\mathcal{X}}^{t}=\left(\chi,\delta u\right)\,,
\label{Xt}
\ee
where $\chi$ and $\delta u$ are the dynamical 
perturbations arising from the gravitational and 
Aether sectors, respectively.
${\bm K}$, ${\bm R}$, ${\bm G}$, and ${\bm M}$ 
are the $2\times2$ symmetric and real matrices, among which only ${\bm M}$ has off-diagonal components. Nonvanishing components of these matrices are 
\ba
\label{eq3.8}
K_{11} &=& \frac{4C_1^2 C_{10}}
{(L-2)(a^2 C_9^2-4C_8 C_{10})},\nn \\
K_{22} &=& \frac{C_1 C_5-C_2^2}{C_1}\,,\nn \\
R_{11} &=& -\frac{a C_9}{C_{10}}K_{11}\,,\qquad 
R_{22}=\frac{C_1 C_6-2C_2 C_3}{C_1}\,,\nn \\
G_{11} &=& \frac{C_8}{C_{10}}K_{11}\,,\qquad 
G_{22}=\frac{C_1 C_7-C_3^2}{C_1}\,,\nn \\
M_{11} &=&-C_1\,,\nn \\
M_{22} &=& L C_{12}+\frac{
L [C_8 C_{11}^2+C_9^2 (C_{10}+a C_{11})]}
{a^2 C_9^2-4C_8 C_{10}}\,, 
\ea
where the explicit form of $M_{12}~(=M_{21})$ is not 
shown here, and
\ba
C_5 &=& \frac{c_1+c_4 a^2 f}{r^2 f}, 
\qquad
C_6 =\frac{2c_4 a b}{r^2}\,,\nn\\
C_7 &=& \frac{[c_4 (a^2 f-1)-c_1]h}{r^2}\,,\nn\\
C_8 &=& -\frac{[c_{13}(a^2 f-1)+1]h}{2r^4}\,,\nn\\
C_9 &=& \frac{c_{13}b}{r^4}\,,\qquad
C_{10} = \frac{1-c_{13}a^2 f}{2r^4 f}, \nn\\
C_{11} &=& \frac{c_{13}a}{r^4}\,, \qquad 
C_{12} =  -\frac{c_1}{r^4}\,.
\ea
Since $M_{12}$ does not depend on $L$, it does not 
affect the angular propagation speeds in 
the large $l$ limit (which will be 
discussed below).

In Ref.~\cite{Tsujikawa:2021typ}, the linear stability conditions of BHs against odd-parity perturbations were derived by using the coordinates $t$ and $r$. As we showed in Sec.~\ref{inssec}, unless a proper coordinate is chosen, 
we may encounter artificial ghosts or Lagrangian 
instabilities in theories with superluminal propagation. 
To overcome this problem, we use the 
Aether-orthogonal frame introduced in Sec.~\ref{Aeorsec}, where 
$\phi$ defines the causality and chronology: 
all particles must move 
along the increasing direction of $\phi$ \cite{Blas:2011ni,Barausse:2011pu}. 
As a result, the future light cone defined by 
each particle with any given speed lies to the future of spacelike hypersurfaces 
($\phi={\rm constant}$), 
as explained explicitly in Ref.~\cite{Kubota:2022lbn}. 

We transform the action (\ref{Sodd}) to that in the 
coordinate system (\ref{phipsi}). We convert the 
$t$ and $r$ derivatives of $\chi$ and $\delta u$ 
to their $\phi$ and $\psi$ derivatives by exploiting 
the relations (\ref{calF1}) and (\ref{calF2}). 
We also use Eq.~(\ref{ab2}) to express $a$ and $b$ 
with respect to $\alpha$. 
Then, the second-order action of odd-parity 
perturbations yields
\be
{\cal S}_{\rm odd}=\sum_{l}L \int 
{\rm d} \phi {\rm d} \psi\,
\hat{{\cal L}}_{\rm odd}\,,
\label{Sodd2}
\ee
where 
\be
\hat{{\cal L}}_{\rm odd}
= \frac{1}{16 \pi G_{\ae}} \sqrt{\frac{h}{f}}
( \vec{\mathcal{X}}^{t}_{,\phi} 
\hat{{\bm K}} \vec{\mathcal{X}}_{,\phi}
+\vec{\mathcal{X}}^{t}_{,\psi}
\hat{{\bm G}} \vec{\mathcal{X}}_{,\psi} 
+\vec{\mathcal{X}}^{t}
\hat{{\bm M}} \vec{\mathcal{X}})\,.
\label{Lodd2}
\ee
Nonvanishing components of the 
$2 \times 2$ matrices 
$\hat{{\bm K}}$, $\hat{{\bm G}}$, and 
$\hat{{\bm M}}$ are given by 
\ba
\hat{K}_{11} &=& \frac{2(1-c_{13})^2 \alpha^2 r^2}
{(L-2)(1+f \alpha^2)^2}\,,\label{K11} \\
\hat{K}_{22} &=& \frac{4c_{14} \alpha^2}
{(1+f \alpha^2)^2} \frac{f}{h}\,,\label{K22} \\
\hat{G}_{11} &=& -\frac{(1+f \alpha^2)^2}{
(1-f \alpha^2)^2} c_T^2 \hat{K}_{11}\,,\\
\hat{G}_{22} &=& -\frac{(1+f \alpha^2)^2}{
(1-f \alpha^2)^2} c_V^2 \hat{K}_{22}\,,\\
\hat{M}_{11} &=& -\frac{1}{2} \left( 1-c_{13} 
\right)\,,\\
\hat{M}_{22} &=& -L\frac{2c_1 -c_{13} (2c_1-c_{13})}
{2(1-c_{13})r^2} \frac{f}{h}\,,
\ea
besides the off-diagonal components 
$\hat{M}_{12}=\hat{M}_{21}$ (which are of order $L^0$).  
The quantities $c_T^2$ and $c_V^2$ are defined, 
respectively, by Eqs.~(\ref{cTm}) and (\ref{cVm}).
We recall that the Lagrangian (\ref{Lodd}) possesses
products of the $t$ and $r$ derivatives, but 
the Lagrangian (\ref{Lodd2}) does not contain  
products of the $\phi$ and $\psi$ derivatives.

The absence of ghosts for dynamical perturbations 
$\chi$ and $\delta u$ requires the two conditions 
$\hat{K}_{11}>0$ and $\hat{K}_{22}>0$. 
The former is satisfied except for 
$c_{13}=1$ (in which case $\hat{K}_{11}=0$). 
The second holds under the inequality 
\be
c_{14}>0\,.
\label{nogo}
\ee
This is equivalent to the no-ghost condition 
of vector perturbations on the Minkowski 
background \cite{Jacobson:2004ts,Oost:2018tcv}. 
On the other hand, it does not coincide with 
the no-ghost condition derived on the SSS background (\ref{SSS}) with the coordinates $t$ 
and $r$ \cite{Tsujikawa:2021typ}. 
The latter coordinate choice is not suitable for 
discussing the linear stability of BHs, since 
the constant $t$ hypersurfaces are not always spacelike. 

To study the propagation of small-scale perturbations with large angular frequencies $\omega$ and momenta $k$, we assume the solutions 
to the perturbation equations for $\chi$ and 
$\delta u$ in the form
\be
\vec{\mathcal{X}}=\vec{\mathcal{X}}_0 
e^{-i(\omega \phi - k \psi)}\,,
\label{calX}
\ee
with $\vec{\mathcal{X}}_0=(\chi_0,\delta u_0)$, where 
$\chi_0$ and $\delta u_0$ are constants.

The radial propagation speeds can be known by considering 
the modes with $\omega r_{\rm g} \approx 
k r_{\rm g} \gg l \gg 1$. In this regime, 
we substitute Eq.~(\ref{calX}) into the perturbation 
equations following from Eq.~(\ref{Lodd2}). 
This leads to the two dispersion relations 
\ba
\omega^2 &=&
-\frac{\hat{G}_{11}}{\hat{K}_{11}}k^2
=\frac{(1+f \alpha^2)^2}{(1-f \alpha^2)^2} c_T^2 k^2\,,
\label{ome1}\\
\omega^2 &=&-\frac{\hat{G}_{22}}{\hat{K}_{22}}k^2
=\frac{(1+f \alpha^2)^2}{(1-f \alpha^2)^2} c_V^2 k^2\,,
\label{ome2}
\ea
which correspond to those of $\chi$ 
and $\delta u$, respectively.  Considering a given point $P$, in the neighborhood of which
we can always express the line element (\ref{phipsi}) 
in the form 
\be
\rd s^2=-\rd \tilde{\phi}^2
+\rd \tilde{\psi}^2+r^2 \Omega_{pq} 
\rd \vartheta^p \rd\vartheta^q\,,
\label{tphi}
\ee
where 
\be
\rd \tilde{\phi}^2=\frac{(1+f\alpha^2)^2}
{4\alpha^2}\rd \phi^2\,,\qquad
\rd \tilde{\psi}^2=\frac{(1-f\alpha^2)^2}
{4\alpha^2}\rd \psi^2\,.
\ee
Note that $\tilde{\phi}$ corresponds to a proper 
time for this coordinate. 
Then, the radial propagation speed squared yields
\be
c_r^2=\left( \frac{\rd \tilde{\psi}}{\rd \tilde{\phi}} 
\right)^2=\frac{(1-f \alpha^2)^2}{(1+f \alpha^2)^2}
\left( \frac{\rd \psi}{\rd \phi} \right)^2
=\frac{(1-f \alpha^2)^2}{(1+f \alpha^2)^2}
\frac{\omega^2}{k^2},
\ee
where we used $(\rd \psi/\rd \phi)^2=\omega^2/k^2$. 
Then, from Eqs.~(\ref{ome1}) and (\ref{ome2}), the 
radial squared propagation speeds of $\chi$ and 
$\delta u$ are given, respectively, by 
\ba
& &
c_{r1}^2=c_T^2=\frac{1}{1-c_{13}}\,,
\label{cr1}\\
& &
c_{r2}^2=c_V^2=\frac{2c_1 -c_{13} (2c_1-c_{13})}
{2c_{14}(1-c_{13})}\,.
\label{cr2}
\ea
Thus, they are identical to the squared 
tensor and vector propagation speeds on the Minkowski background, respectively. 
These values are different from 
those derived on the SSS background (\ref{SSS}) 
with the coordinates $t$ and $r$ \cite{Tsujikawa:2021typ}. 
To avoid the Laplacian instabilities along the radial 
direction, we require the two conditions $c_T^2>0$ 
and $c_V^2>0$. Under the no-ghost condition (\ref{nogo}), 
they amount to the inequalities 
\ba
& &
c_{13}<1\,,\label{Lap1} \\
& &
2c_1 -c_{13} (2c_1-c_{13})>0\,.
\label{Lap2} 
\ea
Under the inequality (\ref{Lap1}), 
the other no-ghost condition $\hat{K}_{11}>0$ is also satisfied.

To derive the angular propagation speeds, we consider the 
eikonal limit $l \approx \omega r_{\rm g} 
\gg k r_{\rm g} \gg 1$. 
We substitute the solution (\ref{calX}) into 
the perturbation equations of motion by noting that 
the off-diagonal matrix components 
$\hat{M}_{12}=\hat{M}_{21}$ are of order $L^0$. 
Then, in the eikonal limit, we obtain 
\ba
\hspace{-0.8cm}
\omega^2 &=&
-\frac{\hat{M}_{11}}{\hat{K}_{11}}
=\frac{(L-2)(1+f \alpha^2)^2}{4(1-c_{13}) \alpha^2 r^2}\,,
\label{ome3}\\
\hspace{-0.8cm}
\omega^2 &=&-\frac{\hat{M}_{22}}{\hat{K}_{22}}
=\frac{L(1+f \alpha^2)^2[2c_1-c_{13}(2c_1-c_{13})]}
{8c_{14}(1-c_{13}) \alpha^2 r^2} \,,
\label{ome4}
\ea
which correspond to the dispersion relations  of $\chi$ and $\delta u$, respectively.
In terms of the proper time $\tilde{\phi}$ in the 
coordinate (\ref{tphi}), the propagation speed squared 
in the $\theta$ direction is given by 
\ba
c_{\Omega}^2&=&\left( \frac{r \rd \theta}
{\rd \tilde{\phi}} \right)^2
=\frac{4\alpha^2}{(1+f \alpha^2)^2}
\left( \frac{r \rd \theta}
{\rd \phi} \right)^2\nonumber\\
&=&\frac{4\alpha^2}{(1+f \alpha^2)^2}
\frac{r^2 \omega^2}{l^2},
\label{cOme}
\ea
where we used $\rd \theta/\rd \phi=\omega/l$.
Substituting Eqs.~(\ref{ome3}) and (\ref{ome4}) into 
Eq.~(\ref{cOme}) and taking the limit $l \gg 1$, 
the squared propagation speeds of $\chi$ and $\delta u$ 
are given, respectively, by 
\ba
& &
c_{\Omega 1}^2=\frac{1}{1-c_{13}}=c_T^2\,,\\
& &
c_{\Omega 2}^2=\frac{2c_1 -c_{13} (2c_1-c_{13})}
{2c_{14}(1-c_{13})}=c_V^2\,.
\ea
These values are equivalent to $c_{r1}^2$ and $c_{r2}^2$
derived in Eqs.~(\ref{cr1}) and (\ref{cr2}), respectively.
Thus, the perturbations $\chi$ and $\delta u$ propagate 
with the same sound speeds as those in the Minkowski 
spacetime both along the radial and angular directions. 
The linear stability of BHs is ensured under the three 
conditions (\ref{nogo}), (\ref{Lap1}), and (\ref{Lap2}).

\subsection{$l=1$}

For the dipole mode ($l=1$ and $L=2$), the metric components $h_{ab}$ 
vanish identically and hence there is a residual gauge degree of freedom to be fixed. 
We choose the gauge $W=0$ and introduce the Lagrangian multiplier 
$\chi$ given by Eq.~(\ref{chidef}). For the coordinate system (\ref{SSS}), the second-order 
action is expressed in the form 
${\cal S}_{\rm odd}=\int \rd t \rd r\,
{\cal L}_{\rm odd}$, where
\begin{widetext}
\ba
{\cal L}_{\rm odd} &=&
\frac{r^2}{8\pi G_{\ae}}\sqrt{\frac{f}{h}} 
\biggl[ C_1 \biggl\{ 2\chi \biggl( -Q'+\frac{2Q}{r}
+\frac{C_2 \dot{\delta u}+C_3 \delta u'+C_4 \delta u}{C_1} 
\biggr)-\chi^2 \biggr\}-\frac{(C_2 \dot{\delta u}+C_3 \delta u'
+C_4 \delta u)^2}{C_1} \nonumber \\
& &\qquad \qquad \quad 
+C_5 \dot{\delta u}^2+C_6 \dot{\delta u} \delta u'
+C_7 \delta u'^2+(2C_{12}+C_{13}) \delta u^2 \biggr]\,,
\label{Loddl=1pre}
\ea
\end{widetext}
with
\ba
C_{13} &=&
\frac{\lambda}{r^2}-\frac{c_{13}[(rh'+2h-2)f+rh f']}
{2r^4 f} \nn\\
& & -\frac{2c_4 (fa'+f'a)ah}{r^3}\,,
\label{C13}
\ea
and $\lambda$ is given in Appendix.
Varying the Lagrangian (\ref{Loddl=1pre}) with respect to $Q$, 
we obtain 
\be
\left( \sqrt{\frac{f}{h}} r^4 C_1 \chi \right)'=0\,.
\label{Qeq}
\ee
We can choose an appropriate boundary condition 
at spatial infinity, such that Eq.~(\ref{Qeq}) 
gives $\chi=0$. 
Then, the Lagrangian (\ref{Loddl=1pre}) reduces to 
\ba
{\cal L}_{\rm odd}
&=&\frac{r^2}{8 \pi G_{\ae}} \sqrt{\frac{f}{h}}
\biggl[ \left( C_5-\frac{C_2^2}{C_1} \right)
{\delta \dot u}^2+ \left(C_7 - \frac{C_3^2}{C_1}\right){\delta u'}^2 \nn\\
&& + \left(C_6 - \frac{2C_2C_3}{C_1}\right) {\delta \dot u} {\delta u}'  
+{\cal M} \delta u^2 \biggr]\,,
\label{Loddl=1}
\ea
where
\be
{\cal M}=2C_{12} +C_{13} -\frac{C_4^2}{C_1}
+\left( \frac{C_3C_4}{C_1} \right)'\,.
\ee
Now, we convert the Lagrangian (\ref{Loddl=1}) 
to that in the Aether-orthogonal frame. 
For this purpose, we replace the derivatives 
${\delta \dot u}$ and $\delta u'$ with $\delta u_{,\phi}$ and $\delta u_{,\psi}$ by using the relations (\ref{calF1}) and (\ref{calF2}). 
Then, the resulting second-order action 
reduces to ${\cal S}_{\rm odd}=\int \rd \phi \rd \psi\,
\hat{\cal L}_{\rm odd}$, where 
\be
\hat{\cal L}_{\rm odd} = \frac{1}{8 \pi G_{\ae}} 
\sqrt{\frac{f}{h}} 
\left( {\cal K} \delta u_{,\phi}^2
+{\cal G} \delta u_{,\psi}^2 +r^2 {\cal M}\delta u^2 \right)\,,
\ee
with
\ba
{\cal K} &=& \frac{4\alpha^2c_{14}}{(1+f \alpha^2)^2}\,,\\
{\cal G} &=& -\frac{2\alpha^2[2c_1 -c_{13} (2c_1-c_{13})]}
{(1-f\alpha^2)^2(1-c_{13})}\,.
\ea
Thus, the Aether perturbation $\delta u$ is the only propagating degree of freedom for $l=1$. 
The ghost is absent under the condition 
${\cal K}>0$, which translates to 
\be
c_{14}>0\,,
\ee
and is the same as Eq.~(\ref{nogo}) derived 
for $l \geq 2$.
The radial squared propagation speed measured 
in terms of the proper time 
$\tilde{\phi}$ reads
\be
c_r^2=-\frac{(1-f \alpha^2)^2}{(1+f \alpha^2)^2}
\frac{{\cal G}}{{\cal K}}=\frac{2c_1 -c_{13} (2c_1-c_{13})}
{2c_{14}(1-c_{13})}=c_V^2\,,
\ee 
which is equivalent to the squared propagation 
speed of vector perturbations  
in the Minkowski spacetime. 
Thus, for $l=1$, there are no additional 
stability conditions to those derived for 
$l \geq 2$.

\subsection{Cases of specific coefficients}
\label{spesec}

For the multiples $l \geq 2$, we consider several specific 
cases in which some of the coefficients 
$c_{1,2,3,4}$ are vanishing. 
{}From Eq.~(\ref{K22}), the matrix component $\hat{K}_{22}$ vanishes for $c_{14}=0$. 
The matrix components $\hat{G}_{22}$ and 
$\hat{M}_{22}$ can be expressed as 
\ba
\hat{G}_{22} &=& -\frac{2[2c_1 -c_{13} (2c_1-c_{13})]\alpha^2}
{(1-c_{13})(1-f \alpha^2)^2} \frac{f}{h}\,,\\
\hat{M}_{22} &=& -\frac{L[2c_1 -c_{13} (2c_1-c_{13})]}
{(1-c_{13})r^2} \frac{f}{h}\,.
\ea
Then, the Aether field either behaves as a vector-type instantaneous mode or exhibits a strong coupling problem for 
\be
c_{14}=0\,,\quad {\rm and} \quad 
2c_1 -c_{13} (2c_1-c_{13}) \neq 0\,,
\label{strong}
\ee
depending on the behavior of the system at 
nonlinear level, the analysis of which is beyond 
the scope of the present paper. 
If we demand that $c_{13}=0$ for the consistency  
with the observational bound (\ref{CD1a}), 
then either a vector-type instantaneous mode or 
a strong coupling problem may arise for 
$c_{14}=0$ and $c_1 \neq 0$, depending on the 
nonlinear behavior of the system. 
This is the case for the stealth Schwarzshild BH 
solution discussed 
in Refs.~\cite{Zhang:2020too,Tsujikawa:2021typ,Zhang:2022fbz}.

If we consider Einstein-Aether theory with 
$c_{14}=0$, $c_1=0$, and $c_{13}=0$, i.e., 
\begin{equation}
c_1=0\,,\qquad c_2 \neq 0\,,\qquad
c_3=0\,,\qquad c_4=0\,,
\label{c2}
\end{equation}
we have the following matrix components
\begin{eqnarray}
& &
\hat{K}_{11}=\frac{2\alpha^2 r^2}{(L-2)(1+f\alpha^2)^2}\,,
\nn \\
& &
\hat{G}_{11}=-\frac{2\alpha^2 r^2}{(L-2)(1-f\alpha^2)^2}\,,
\nn \\
& &
\hat{M}_{11}=-\frac{1}{2}\,,\nn \\
& &
\hat{K}_{22}=\hat{G}_{22}=\hat{M}_{22}=0\,.
\end{eqnarray}
The number of propagating degrees of freedom in the odd-parity sector is 1 at linear level. This indicates either the absence of vector modes, the presence of a vector-type instantaneous mode or the  strong coupling problem for $\delta u$, depending on the nonlinear behavior of this system. Fortunately, in this case, 
we know the nonlinear behavior since Einstein-Aether theory with the coefficients (\ref{c2}) is equivalent to a class of cuscuton theories with a quadratic potential \cite{Bhattacharyya:2016mah}, provided that the derivative of the expansion $\theta=\nabla_{\mu}u^{\mu}$ is 
non-zero\footnote{If $\partial_{\mu}\theta=0$, then $u^{\mu}$ is undetermined by equations of motion and $\lambda=0$.}. This means that the absence of the time kinetic term and the gradient term shown above for the specific coefficients (\ref{c2}) simply corresponds to the absence of vector modes as far as the equivalence to the cuscuton theory holds on the background with $\partial_{\mu}\theta\neq 0$.
There is a single dynamical degree of freedom $\chi$ with the propagation speeds given by 
\begin{equation}
c_{r1}^2=c_{\Omega 1}^2=1\,,
\end{equation}
which are both luminal.

%%%%%%%%%%%%%%%%%%%%%%%%
\section{Conclusions}
\label{consec}
%%%%%%%%%%%%%%%%%%%%%%%%

In this paper, we addressed the linear stability of BHs against 
odd-parity perturbations in Einstein-Aether theory given by 
the action (\ref{action}). In this theory, there is 
a preferred threading aligned with a unit timelike vector field. 
If the background Aether field $u_{\mu}$ has vanishing vorticity, 
one can introduce a scalar (Khronon) field $\phi$ 
whose gradient $\partial_{\mu}\phi$ is timelike and proportional to $u_{\mu}$. 
This property holds for the SSS background given by 
the line element (\ref{SSS}). 

In Einstein-Aether theory, the constant $t$ hypersurfaces in the 
coordinate (\ref{SSS}) are not always spacelike outside the universal horizon, which now is the boundary of 
a BH and is always inside the metric horizon, when superlunimal speeds are allowed \cite{Blas:2011ni}.
In this sense, the derivation of linear stability conditions 
using $t$ as a time clock can lead to inconsistent 
results. The proper coordinate choice for obtaining 
no-ghost conditions and propagation speeds of dynamical perturbations should be the Aether-orthogonal frame 
in which the Khronon field $\phi$ is treated as a time clock, in which case the constant time hypersurfaces are always spacelike over the whole region  outside the universal horizon, as shown explicitly by the metric (\ref{phipsi}).
In Sec.~\ref{inssec}, we argued how the coordinate choice different from the Aether-orthogonal frame can give rise to apparent ghost and Laplacian instabilities.

In Sec.~\ref{stasec}, we derived the second-order action of 
odd-parity perturbations by transforming the action derived 
for the SSS coordinate (\ref{SSS}) in 
Ref.~\cite{Tsujikawa:2021typ} to the one in the 
Aether-orthogonal frame with the line element 
(\ref{phipsi}). For this purpose, we exploited 
transformation properties (\ref{calF1})-(\ref{calF2}) of the 
derivatives of perturbations between the two sets of coordinates.
For the multipoles $l \geq 2$, there are two dynamical perturbations $\chi$ and $\delta u$ arising from the gravitational and vector-field sectors, respectively. 
The resulting second-order Lagrangian is of the form 
(\ref{Lodd2}), which does not contain products of 
the $\phi$ and $\psi$ derivatives (unlike the Lagrangian 
(\ref{Lodd}) containing products of the $t$ and $r$ derivatives).
The stability analysis of BHs in the Aether-orthogonal frame 
shows that the ghost is absent under the inequality $c_{14}>0$, 
which is the same no-ghost condition of vector perturbations 
on the Minkowski background. 
In large momentum limits, the radial squared propagation speeds of 
$\chi$ and $\delta u$ are equivalent to those of the tensor 
and vector perturbations on the Minkowski background. 
This is also the case for the angular squared propagation 
speeds of $\chi$ and $\delta u$ in the eikonal limit $l \gg 1$. 
For $l=1$, the vector-field perturbation alone
propagates with the same stability conditions of $\delta u$ 
as those derived for $l \geq 2$. 

We thus showed that the proper odd-parity stability analysis of 
BHs based on the Aether-orthogonal frame gives rise to the same 
no-ghost conditions and propagation speeds of dynamical 
perturbations as those on the Minkowski background. 
In Sec.~\ref{spesec}, we discussed several specific cases 
of coupling constants in which the strong coupling problem may arise or the number of degrees of freedom reduces. 
It will be of interest to classify surviving BH solutions 
free from the linear instability and strong coupling 
problems. For this purpose, we plan to extend the stability
analysis in the Aether-orthogonal frame to perturbations 
in the even-parity sector.

\section*{Acknowledgements}

The work of SM was supported in part by Japan Society for the Promotion of Science (JSPS) Grants-in-Aid for Scientific Research No.~24K07017 and the World Premier International Research Center Initiative (WPI), MEXT, Japan. 
ST was supported by the Grant-in-Aid for Scientific Research Fund 
of the JSPS No.~22K03642 and Waseda University Special Research 
Project No.~2023C-473. AW is partially supported by a US NSF grant 
with the grant number: PHY2308845.

\section*{Appendix A: The Lagrange multiplier in Spherical Spacetimes}
\renewcommand{\theequation}{A.\arabic{equation}} \setcounter{equation}{0}

The quantity $\lambda$ in Eq.~(\ref{C13}) is given by 
\begin{widetext}
\ba
\label{a2}
\hspace{-0.5cm}
\lambda &=&\frac{1}{4 f^2 r^2 \left(a^2 f-1\right)}
\bigg\{f r^2 \left[a^2 \left(6 c_1+3 c_2+2 c_{13}-8 c_{14}\right) h \left(f'\right)^2+c_2 \left(2 h f''+f' h'\right)\right] \nn\\
\hspace{-0.5cm}
&& +2 a^2 f^4 \left[-2 a^2 \left(2 \left(c_2+c_{13}\right) h-c_2 r h'\right)+4 \left(c_{14}-c_1\right) h r^2 \left(a'\right)^2+a \left(-c_1+c_2+c_{13}\right) r \left(2 h r a''+a' \left(r h'+4 h\right)\right)\right] \nn\\
\hspace{-0.5cm}
&& -f^2 \Big[r h' \left(a^2 \left(-2 c_1+3 c_2+2 c_{13}\right) r f'-4 c_2\right) \nn\\
\hspace{-0.5cm}
&& +2 h \big(a^4 \left(3 c_1+c_2+c_{13}-4 c_{14}\right) r^2 \left(f'\right)^2+a^2 r \left(\left(-2 c_1+3 c_2+2 c_{13}\right) r f''+2 \left(-2 c_1+c_2+2 c_{13}\right) f'\right)\nn\\
\hspace{-0.5cm}
&& -a \left(11 c_1-5 c_2-5 c_{13}-8 c_{14}\right) r^2 a' f'+4 \left(c_2+c_{13}\right)\big)\Big] 
-\left(2 c_2+c_{13}\right) h r^2 \left(f'\right)^2 \nn\\
\hspace{-0.5cm}
&& +2 f^3 \Big[ a^4 r \left(\left(-c_1+c_2+c_{13}\right) r f' h'+2 h \left(\left(-c_1+c_2+c_{13}\right) r f''+\left(-2 c_1+c_2+2 c_{13}\right) f'\right)\right) \nn\\
\hspace{-0.5cm}
&& +a^2 \left(8 \left(c_2+c_{13}\right) h-4 c_2 r h'\right)-2 \left(-2 c_1+c_2+c_{13}+2 c_{14}\right) h r^2 \left(a'\right)^2 \nn\\
\hspace{-0.5cm}
&& +a^3 \left(-11 c_1+3 c_2+3 c_{13}+8 c_{14}\right) h r^2 a' f'+a \left(c_1-c_2-c_{13}\right) r \left(2 h r a''+a' \left(r h'+4 h\right)\right)\Big] \bigg\}.~~~
\ea
\end{widetext}

\bibliographystyle{mybibstyle}
\bibliography{bib}

\begin{thebibliography}{78}%
\makeatletter
\providecommand \@ifxundefined [1]{%
 \@ifx{#1\undefined}
}%
\providecommand \@ifnum [1]{%
 \ifnum #1\expandafter \@firstoftwo
 \else \expandafter \@secondoftwo
 \fi
}%
\providecommand \@ifx [1]{%
 \ifx #1\expandafter \@firstoftwo
 \else \expandafter \@secondoftwo
 \fi
}%
\providecommand \natexlab [1]{#1}%
\providecommand \enquote  [1]{``#1''}%
\providecommand \bibnamefont  [1]{#1}%
\providecommand \bibfnamefont [1]{#1}%
\providecommand \citenamefont [1]{#1}%
\providecommand \href@noop [0]{\@secondoftwo}%
\providecommand \href [0]{\begingroup \@sanitize@url \@href}%
\providecommand \@href[1]{\@@startlink{#1}\@@href}%
\providecommand \@@href[1]{\endgroup#1\@@endlink}%
\providecommand \@sanitize@url [0]{\catcode `\\12\catcode `\$12\catcode
  `\&12\catcode `\#12\catcode `\^12\catcode `\_12\catcode `\%12\relax}%
\providecommand \@@startlink[1]{}%
\providecommand \@@endlink[0]{}%
\providecommand \url  [0]{\begingroup\@sanitize@url \@url }%
\providecommand \@url [1]{\endgroup\@href {#1}{\urlprefix }}%
\providecommand \urlprefix  [0]{URL }%
\providecommand \Eprint [0]{\href }%
\providecommand \doibase [0]{http://dx.doi.org/}%
\providecommand \selectlanguage [0]{\@gobble}%
\providecommand \bibinfo  [0]{\@secondoftwo}%
\providecommand \bibfield  [0]{\@secondoftwo}%
\providecommand \translation [1]{[#1]}%
\providecommand \BibitemOpen [0]{}%
\providecommand \bibitemStop [0]{}%
\providecommand \bibitemNoStop [0]{.\EOS\space}%
\providecommand \EOS [0]{\spacefactor3000\relax}%
\providecommand \BibitemShut  [1]{\csname bibitem#1\endcsname}%
\let\auto@bib@innerbib\@empty
%</preamble>
\bibitem [{\citenamefont {Kostelecky}\ and\ \citenamefont
  {Samuel}(1989)}]{Kostelecky:1988zi}%
  \BibitemOpen
  \bibfield  {author} {\bibinfo {author} {\bibfnamefont {V.~A.}\ \bibnamefont
  {Kostelecky}} and \bibinfo {author} {\bibfnamefont {S.}~\bibnamefont
  {Samuel}},\ }\href {\doibase 10.1103/PhysRevD.39.683} {\bibfield  {journal}
  {\bibinfo  {journal} {\emph {Phys. Rev. D}}\ }\textbf {\bibinfo {volume}
  {39}},\ \bibinfo {pages} {683} (\bibinfo {year} {1989})}\BibitemShut
  {NoStop}%
\bibitem [{\citenamefont {Gambini}\ and\ \citenamefont
  {Pullin}(1999)}]{Gambini:1998it}%
  \BibitemOpen
  \bibfield  {author} {\bibinfo {author} {\bibfnamefont {R.}~\bibnamefont
  {Gambini}} and \bibinfo {author} {\bibfnamefont {J.}~\bibnamefont {Pullin}},\
  }\href {\doibase 10.1103/PhysRevD.59.124021} {\bibfield  {journal} {\bibinfo
  {journal} {\emph {Phys. Rev. D}}\ }\textbf {\bibinfo {volume} {59}},\
  \bibinfo {pages} {124021} (\bibinfo {year} {1999})},\ \Eprint
  {http://arxiv.org/abs/gr-qc/9809038} {arXiv:gr-qc/9809038} \BibitemShut
  {NoStop}%
\bibitem [{\citenamefont {Douglas}\ and\ \citenamefont
  {Nekrasov}(2001)}]{Douglas:2001ba}%
  \BibitemOpen
  \bibfield  {author} {\bibinfo {author} {\bibfnamefont {M.~R.}\ \bibnamefont
  {Douglas}} and \bibinfo {author} {\bibfnamefont {N.~A.}\ \bibnamefont
  {Nekrasov}},\ }\href {\doibase 10.1103/RevModPhys.73.977} {\bibfield
  {journal} {\bibinfo  {journal} {\emph {Rev. Mod. Phys.}}\ }\textbf {\bibinfo
  {volume} {73}},\ \bibinfo {pages} {977} (\bibinfo {year} {2001})},\ \Eprint
  {http://arxiv.org/abs/hep-th/0106048} {arXiv:hep-th/0106048} \BibitemShut
  {NoStop}%
\bibitem [{\citenamefont {Carroll}\ \emph {et~al.}(2001)\citenamefont
  {Carroll}, \citenamefont {Harvey}, \citenamefont {Kostelecky}, \citenamefont
  {Lane},\ and\ \citenamefont {Okamoto}}]{Carroll:2001ws}%
  \BibitemOpen
  \bibfield  {author} {\bibinfo {author} {\bibfnamefont {S.~M.}\ \bibnamefont
  {Carroll}}, \bibinfo {author} {\bibfnamefont {J.~A.}\ \bibnamefont {Harvey}},
  \bibinfo {author} {\bibfnamefont {V.~A.}\ \bibnamefont {Kostelecky}},
  \bibinfo {author} {\bibfnamefont {C.~D.}\ \bibnamefont {Lane}},  and \bibinfo
  {author} {\bibfnamefont {T.}~\bibnamefont {Okamoto}},\ }\href {\doibase
  10.1103/PhysRevLett.87.141601} {\bibfield  {journal} {\bibinfo  {journal}
  {\emph {Phys. Rev. Lett.}}\ }\textbf {\bibinfo {volume} {87}},\ \bibinfo
  {pages} {141601} (\bibinfo {year} {2001})},\ \Eprint
  {http://arxiv.org/abs/hep-th/0105082} {arXiv:hep-th/0105082} \BibitemShut
  {NoStop}%
\bibitem [{\citenamefont {Amelino-Camelia}(2013)}]{Amelino-Camelia:2008aez}%
  \BibitemOpen
  \bibfield  {author} {\bibinfo {author} {\bibfnamefont {G.}~\bibnamefont
  {Amelino-Camelia}},\ }\href {\doibase 10.12942/lrr-2013-5} {\bibfield
  {journal} {\bibinfo  {journal} {\emph {Living Rev. Rel.}}\ }\textbf {\bibinfo
  {volume} {16}},\ \bibinfo {pages} {5} (\bibinfo {year} {2013})},\ \Eprint
  {http://arxiv.org/abs/0806.0339} {arXiv:0806.0339 [gr-qc]} \BibitemShut
  {NoStop}%
\bibitem [{\citenamefont {Chadha}\ and\ \citenamefont
  {Nielsen}(1983)}]{Chadha:1982qq}%
  \BibitemOpen
  \bibfield  {author} {\bibinfo {author} {\bibfnamefont {S.}~\bibnamefont
  {Chadha}} and \bibinfo {author} {\bibfnamefont {H.~B.}\ \bibnamefont
  {Nielsen}},\ }\href {\doibase 10.1016/0550-3213(83)90081-0} {\bibfield
  {journal} {\bibinfo  {journal} {\emph {Nucl. Phys. B}}\ }\textbf {\bibinfo
  {volume} {217}},\ \bibinfo {pages} {125} (\bibinfo {year}
  {1983})}\BibitemShut {NoStop}%
\bibitem [{\citenamefont {Mattingly}(2005)}]{Mattingly:2005re}%
  \BibitemOpen
  \bibfield  {author} {\bibinfo {author} {\bibfnamefont {D.}~\bibnamefont
  {Mattingly}},\ }\href {\doibase 10.12942/lrr-2005-5} {\bibfield  {journal}
  {\bibinfo  {journal} {\emph {Living Rev. Rel.}}\ }\textbf {\bibinfo {volume}
  {8}},\ \bibinfo {pages} {5} (\bibinfo {year} {2005})},\ \Eprint
  {http://arxiv.org/abs/gr-qc/0502097} {arXiv:gr-qc/0502097} \BibitemShut
  {NoStop}%
\bibitem [{\citenamefont {Kostelecky}\ and\ \citenamefont
  {Russell}(2011)}]{Kostelecky:2008ts}%
  \BibitemOpen
  \bibfield  {author} {\bibinfo {author} {\bibfnamefont {V.~A.}\ \bibnamefont
  {Kostelecky}} and \bibinfo {author} {\bibfnamefont {N.}~\bibnamefont
  {Russell}},\ }\href {\doibase 10.1103/RevModPhys.83.11} {\bibfield  {journal}
  {\bibinfo  {journal} {\emph {Rev. Mod. Phys.}}\ }\textbf {\bibinfo {volume}
  {83}},\ \bibinfo {pages} {11} (\bibinfo {year} {2011})},\ \Eprint
  {http://arxiv.org/abs/0801.0287} {arXiv:0801.0287 [hep-ph]} \BibitemShut
  {NoStop}%
\bibitem [{\citenamefont {Will}(2014)}]{Will:2014kxa}%
  \BibitemOpen
  \bibfield  {author} {\bibinfo {author} {\bibfnamefont {C.~M.}\ \bibnamefont
  {Will}},\ }\href {\doibase 10.12942/lrr-2014-4} {\bibfield  {journal}
  {\bibinfo  {journal} {\emph {Living Rev. Rel.}}\ }\textbf {\bibinfo {volume}
  {17}},\ \bibinfo {pages} {4} (\bibinfo {year} {2014})},\ \Eprint
  {http://arxiv.org/abs/1403.7377} {arXiv:1403.7377 [gr-qc]} \BibitemShut
  {NoStop}%
\bibitem [{\citenamefont {Jacobson}\ and\ \citenamefont
  {Mattingly}(2001)}]{Jacobson:2000xp}%
  \BibitemOpen
  \bibfield  {author} {\bibinfo {author} {\bibfnamefont {T.}~\bibnamefont
  {Jacobson}} and \bibinfo {author} {\bibfnamefont {D.}~\bibnamefont
  {Mattingly}},\ }\href {\doibase 10.1103/PhysRevD.64.024028} {\bibfield
  {journal} {\bibinfo  {journal} {\emph {Phys. Rev. D}}\ }\textbf {\bibinfo
  {volume} {64}},\ \bibinfo {pages} {024028} (\bibinfo {year} {2001})},\
  \Eprint {http://arxiv.org/abs/gr-qc/0007031} {arXiv:gr-qc/0007031}
  \BibitemShut {NoStop}%
\bibitem [{\citenamefont {Gasperini}(1987)}]{Gasperini:1987nq}%
  \BibitemOpen
  \bibfield  {author} {\bibinfo {author} {\bibfnamefont {M.}~\bibnamefont
  {Gasperini}},\ }\href {\doibase 10.1088/0264-9381/4/2/026} {\bibfield
  {journal} {\bibinfo  {journal} {\emph {Class. Quant. Grav.}}\ }\textbf
  {\bibinfo {volume} {4}},\ \bibinfo {pages} {485} (\bibinfo {year}
  {1987})}\BibitemShut {NoStop}%
\bibitem [{\citenamefont {Afshordi}\ \emph {et~al.}(2007)\citenamefont
  {Afshordi}, \citenamefont {Chung},\ and\ \citenamefont
  {Geshnizjani}}]{Afshordi:2006ad}%
  \BibitemOpen
  \bibfield  {author} {\bibinfo {author} {\bibfnamefont {N.}~\bibnamefont
  {Afshordi}}, \bibinfo {author} {\bibfnamefont {D.~J.~H.}\ \bibnamefont
  {Chung}},  and \bibinfo {author} {\bibfnamefont {G.}~\bibnamefont
  {Geshnizjani}},\ }\href {\doibase 10.1103/PhysRevD.75.083513} {\bibfield
  {journal} {\bibinfo  {journal} {\emph {Phys. Rev. D}}\ }\textbf {\bibinfo
  {volume} {75}},\ \bibinfo {pages} {083513} (\bibinfo {year} {2007})},\
  \Eprint {http://arxiv.org/abs/hep-th/0609150} {arXiv:hep-th/0609150}
  \BibitemShut {NoStop}%
\bibitem [{\citenamefont {Bhattacharyya}\ \emph {et~al.}(2018)\citenamefont
  {Bhattacharyya}, \citenamefont {Coates}, \citenamefont {Colombo},
  \citenamefont {Gumrukcuoglu},\ and\ \citenamefont
  {Sotiriou}}]{Bhattacharyya:2016mah}%
  \BibitemOpen
  \bibfield  {author} {\bibinfo {author} {\bibfnamefont {J.}~\bibnamefont
  {Bhattacharyya}}, \bibinfo {author} {\bibfnamefont {A.}~\bibnamefont
  {Coates}}, \bibinfo {author} {\bibfnamefont {M.}~\bibnamefont {Colombo}},
  \bibinfo {author} {\bibfnamefont {A.~E.}\ \bibnamefont {Gumrukcuoglu}},  and
  \bibinfo {author} {\bibfnamefont {T.~P.}\ \bibnamefont {Sotiriou}},\ }\href
  {\doibase 10.1103/PhysRevD.97.064020} {\bibfield  {journal} {\bibinfo
  {journal} {\emph {Phys. Rev. D}}\ }\textbf {\bibinfo {volume} {97}},\
  \bibinfo {pages} {064020} (\bibinfo {year} {2018})},\ \Eprint
  {http://arxiv.org/abs/1612.01824} {arXiv:1612.01824 [hep-th]} \BibitemShut
  {NoStop}%
\bibitem [{\citenamefont {Gripaios}(2004)}]{Gripaios:2004ms}%
  \BibitemOpen
  \bibfield  {author} {\bibinfo {author} {\bibfnamefont {B.~M.}\ \bibnamefont
  {Gripaios}},\ }\href {\doibase 10.1088/1126-6708/2004/10/069} {\bibfield
  {journal} {\bibinfo  {journal} {\emph {JHEP}}\ }\textbf {\bibinfo {volume}
  {10}},\ \bibinfo {pages} {069} (\bibinfo {year} {2004})},\ \Eprint
  {http://arxiv.org/abs/hep-th/0408127} {arXiv:hep-th/0408127} \BibitemShut
  {NoStop}%
\bibitem [{\citenamefont {Zlosnik}\ \emph {et~al.}(2007)\citenamefont
  {Zlosnik}, \citenamefont {Ferreira},\ and\ \citenamefont
  {Starkman}}]{Zlosnik:2006zu}%
  \BibitemOpen
  \bibfield  {author} {\bibinfo {author} {\bibfnamefont {T.~G.}\ \bibnamefont
  {Zlosnik}}, \bibinfo {author} {\bibfnamefont {P.~G.}\ \bibnamefont
  {Ferreira}},  and \bibinfo {author} {\bibfnamefont {G.~D.}\ \bibnamefont
  {Starkman}},\ }\href {\doibase 10.1103/PhysRevD.75.044017} {\bibfield
  {journal} {\bibinfo  {journal} {\emph {Phys. Rev. D}}\ }\textbf {\bibinfo
  {volume} {75}},\ \bibinfo {pages} {044017} (\bibinfo {year} {2007})},\
  \Eprint {http://arxiv.org/abs/astro-ph/0607411} {arXiv:astro-ph/0607411}
  \BibitemShut {NoStop}%
\bibitem [{\citenamefont {Kanno}\ and\ \citenamefont
  {Soda}(2006)}]{Kanno:2006ty}%
  \BibitemOpen
  \bibfield  {author} {\bibinfo {author} {\bibfnamefont {S.}~\bibnamefont
  {Kanno}} and \bibinfo {author} {\bibfnamefont {J.}~\bibnamefont {Soda}},\
  }\href {\doibase 10.1103/PhysRevD.74.063505} {\bibfield  {journal} {\bibinfo
  {journal} {\emph {Phys. Rev. D}}\ }\textbf {\bibinfo {volume} {74}},\
  \bibinfo {pages} {063505} (\bibinfo {year} {2006})},\ \Eprint
  {http://arxiv.org/abs/hep-th/0604192} {arXiv:hep-th/0604192} \BibitemShut
  {NoStop}%
\bibitem [{\citenamefont {Chesler}\ and\ \citenamefont
  {Loeb}(2017)}]{Chesler:2017khz}%
  \BibitemOpen
  \bibfield  {author} {\bibinfo {author} {\bibfnamefont {P.~M.}\ \bibnamefont
  {Chesler}} and \bibinfo {author} {\bibfnamefont {A.}~\bibnamefont {Loeb}},\
  }\href {\doibase 10.1103/PhysRevLett.119.031102} {\bibfield  {journal}
  {\bibinfo  {journal} {\emph {Phys. Rev. Lett.}}\ }\textbf {\bibinfo {volume}
  {119}},\ \bibinfo {pages} {031102} (\bibinfo {year} {2017})},\ \Eprint
  {http://arxiv.org/abs/1704.05116} {arXiv:1704.05116 [astro-ph.HE]}
  \BibitemShut {NoStop}%
\bibitem [{\citenamefont {Jacobson}\ and\ \citenamefont
  {Mattingly}(2004)}]{Jacobson:2004ts}%
  \BibitemOpen
  \bibfield  {author} {\bibinfo {author} {\bibfnamefont {T.}~\bibnamefont
  {Jacobson}} and \bibinfo {author} {\bibfnamefont {D.}~\bibnamefont
  {Mattingly}},\ }\href {\doibase 10.1103/PhysRevD.70.024003} {\bibfield
  {journal} {\bibinfo  {journal} {\emph {Phys. Rev. D}}\ }\textbf {\bibinfo
  {volume} {70}},\ \bibinfo {pages} {024003} (\bibinfo {year} {2004})},\
  \Eprint {http://arxiv.org/abs/gr-qc/0402005} {arXiv:gr-qc/0402005}
  \BibitemShut {NoStop}%
\bibitem [{\citenamefont {Abbott}\ \emph {et~al.}(2017)\citenamefont {Abbott}
  \emph {et~al.}}]{LIGOScientific:2017zic}%
  \BibitemOpen
  \bibfield  {author} {\bibinfo {author} {\bibfnamefont {B.~P.}\ \bibnamefont
  {Abbott}} \emph {et~al.} (\bibinfo {collaboration} {LIGO Scientific, Virgo,
  Fermi-GBM, INTEGRAL}),\ }\href {\doibase 10.3847/2041-8213/aa920c} {\bibfield
   {journal} {\bibinfo  {journal} {\emph {Astrophys. J. Lett.}}\ }\textbf
  {\bibinfo {volume} {848}},\ \bibinfo {pages} {L13} (\bibinfo {year}
  {2017})},\ \Eprint {http://arxiv.org/abs/1710.05834} {arXiv:1710.05834
  [astro-ph.HE]} \BibitemShut {NoStop}%
\bibitem [{\citenamefont {Gong}\ \emph {et~al.}(2018)\citenamefont {Gong},
  \citenamefont {Hou}, \citenamefont {Liang},\ and\ \citenamefont
  {Papantonopoulos}}]{Gong:2018cgj}%
  \BibitemOpen
  \bibfield  {author} {\bibinfo {author} {\bibfnamefont {Y.}~\bibnamefont
  {Gong}}, \bibinfo {author} {\bibfnamefont {S.}~\bibnamefont {Hou}}, \bibinfo
  {author} {\bibfnamefont {D.}~\bibnamefont {Liang}},  and \bibinfo {author}
  {\bibfnamefont {E.}~\bibnamefont {Papantonopoulos}},\ }\href {\doibase
  10.1103/PhysRevD.97.084040} {\bibfield  {journal} {\bibinfo  {journal} {\emph
  {Phys. Rev. D}}\ }\textbf {\bibinfo {volume} {97}},\ \bibinfo {pages}
  {084040} (\bibinfo {year} {2018})},\ \Eprint
  {http://arxiv.org/abs/1801.03382} {arXiv:1801.03382 [gr-qc]} \BibitemShut
  {NoStop}%
\bibitem [{\citenamefont {Oost}\ \emph {et~al.}(2018)\citenamefont {Oost},
  \citenamefont {Mukohyama},\ and\ \citenamefont {Wang}}]{Oost:2018tcv}%
  \BibitemOpen
  \bibfield  {author} {\bibinfo {author} {\bibfnamefont {J.}~\bibnamefont
  {Oost}}, \bibinfo {author} {\bibfnamefont {S.}~\bibnamefont {Mukohyama}},
  and \bibinfo {author} {\bibfnamefont {A.}~\bibnamefont {Wang}},\ }\href
  {\doibase 10.1103/PhysRevD.97.124023} {\bibfield  {journal} {\bibinfo
  {journal} {\emph {Phys. Rev. D}}\ }\textbf {\bibinfo {volume} {97}},\
  \bibinfo {pages} {124023} (\bibinfo {year} {2018})},\ \Eprint
  {http://arxiv.org/abs/1802.04303} {arXiv:1802.04303 [gr-qc]} \BibitemShut
  {NoStop}%
\bibitem [{\citenamefont {Elliott}\ \emph {et~al.}(2005)\citenamefont
  {Elliott}, \citenamefont {Moore},\ and\ \citenamefont
  {Stoica}}]{Elliott:2005va}%
  \BibitemOpen
  \bibfield  {author} {\bibinfo {author} {\bibfnamefont {J.~W.}\ \bibnamefont
  {Elliott}}, \bibinfo {author} {\bibfnamefont {G.~D.}\ \bibnamefont {Moore}},
  and \bibinfo {author} {\bibfnamefont {H.}~\bibnamefont {Stoica}},\ }\href
  {\doibase 10.1088/1126-6708/2005/08/066} {\bibfield  {journal} {\bibinfo
  {journal} {\emph {JHEP}}\ }\textbf {\bibinfo {volume} {08}},\ \bibinfo
  {pages} {066} (\bibinfo {year} {2005})},\ \Eprint
  {http://arxiv.org/abs/hep-ph/0505211} {arXiv:hep-ph/0505211} \BibitemShut
  {NoStop}%
\bibitem [{\citenamefont {Carroll}\ and\ \citenamefont
  {Lim}(2004)}]{Carroll:2004ai}%
  \BibitemOpen
  \bibfield  {author} {\bibinfo {author} {\bibfnamefont {S.~M.}\ \bibnamefont
  {Carroll}} and \bibinfo {author} {\bibfnamefont {E.~A.}\ \bibnamefont
  {Lim}},\ }\href {\doibase 10.1103/PhysRevD.70.123525} {\bibfield  {journal}
  {\bibinfo  {journal} {\emph {Phys. Rev. D}}\ }\textbf {\bibinfo {volume}
  {70}},\ \bibinfo {pages} {123525} (\bibinfo {year} {2004})},\ \Eprint
  {http://arxiv.org/abs/hep-th/0407149} {arXiv:hep-th/0407149} \BibitemShut
  {NoStop}%
\bibitem [{\citenamefont {Foster}\ and\ \citenamefont
  {Jacobson}(2006)}]{Foster:2005dk}%
  \BibitemOpen
  \bibfield  {author} {\bibinfo {author} {\bibfnamefont {B.~Z.}\ \bibnamefont
  {Foster}} and \bibinfo {author} {\bibfnamefont {T.}~\bibnamefont
  {Jacobson}},\ }\href {\doibase 10.1103/PhysRevD.73.064015} {\bibfield
  {journal} {\bibinfo  {journal} {\emph {Phys. Rev. D}}\ }\textbf {\bibinfo
  {volume} {73}},\ \bibinfo {pages} {064015} (\bibinfo {year} {2006})},\
  \Eprint {http://arxiv.org/abs/gr-qc/0509083} {arXiv:gr-qc/0509083}
  \BibitemShut {NoStop}%
\bibitem [{\citenamefont {Foster}(2007)}]{Foster:2007gr}%
  \BibitemOpen
  \bibfield  {author} {\bibinfo {author} {\bibfnamefont {B.~Z.}\ \bibnamefont
  {Foster}},\ }\href {\doibase 10.1103/PhysRevD.76.084033} {\bibfield
  {journal} {\bibinfo  {journal} {\emph {Phys. Rev. D}}\ }\textbf {\bibinfo
  {volume} {76}},\ \bibinfo {pages} {084033} (\bibinfo {year} {2007})},\
  \Eprint {http://arxiv.org/abs/0706.0704} {arXiv:0706.0704 [gr-qc]}
  \BibitemShut {NoStop}%
\bibitem [{\citenamefont {Yagi}\ \emph
  {et~al.}(2014{\natexlab{a}})\citenamefont {Yagi}, \citenamefont {Blas},
  \citenamefont {Yunes},\ and\ \citenamefont {Barausse}}]{Yagi:2013qpa}%
  \BibitemOpen
  \bibfield  {author} {\bibinfo {author} {\bibfnamefont {K.}~\bibnamefont
  {Yagi}}, \bibinfo {author} {\bibfnamefont {D.}~\bibnamefont {Blas}}, \bibinfo
  {author} {\bibfnamefont {N.}~\bibnamefont {Yunes}},  and \bibinfo {author}
  {\bibfnamefont {E.}~\bibnamefont {Barausse}},\ }\href {\doibase
  10.1103/PhysRevLett.112.161101} {\bibfield  {journal} {\bibinfo  {journal}
  {\emph {Phys. Rev. Lett.}}\ }\textbf {\bibinfo {volume} {112}},\ \bibinfo
  {pages} {161101} (\bibinfo {year} {2014}{\natexlab{a}})},\ \Eprint
  {http://arxiv.org/abs/1307.6219} {arXiv:1307.6219 [gr-qc]} \BibitemShut
  {NoStop}%
\bibitem [{\citenamefont {Yagi}\ \emph
  {et~al.}(2014{\natexlab{b}})\citenamefont {Yagi}, \citenamefont {Blas},
  \citenamefont {Barausse},\ and\ \citenamefont {Yunes}}]{Yagi:2013ava}%
  \BibitemOpen
  \bibfield  {author} {\bibinfo {author} {\bibfnamefont {K.}~\bibnamefont
  {Yagi}}, \bibinfo {author} {\bibfnamefont {D.}~\bibnamefont {Blas}}, \bibinfo
  {author} {\bibfnamefont {E.}~\bibnamefont {Barausse}},  and \bibinfo {author}
  {\bibfnamefont {N.}~\bibnamefont {Yunes}},\ }\href {\doibase
  10.1103/PhysRevD.89.084067} {\bibfield  {journal} {\bibinfo  {journal} {\emph
  {Phys. Rev. D}}\ }\textbf {\bibinfo {volume} {89}},\ \bibinfo {pages}
  {084067} (\bibinfo {year} {2014}{\natexlab{b}})},\ \bibinfo {note} {[Erratum:
  Phys.Rev.D 90, 069902 (2014), Erratum: Phys.Rev.D 90, 069901 (2014)]},\
  \Eprint {http://arxiv.org/abs/1311.7144} {arXiv:1311.7144 [gr-qc]}
  \BibitemShut {NoStop}%
\bibitem [{\citenamefont {Gupta}\ \emph {et~al.}(2021)\citenamefont {Gupta},
  \citenamefont {Herrero-Valea}, \citenamefont {Blas}, \citenamefont
  {Barausse}, \citenamefont {Cornish}, \citenamefont {Yagi},\ and\
  \citenamefont {Yunes}}]{Gupta:2021vdj}%
  \BibitemOpen
  \bibfield  {author} {\bibinfo {author} {\bibfnamefont {T.}~\bibnamefont
  {Gupta}}, \bibinfo {author} {\bibfnamefont {M.}~\bibnamefont
  {Herrero-Valea}}, \bibinfo {author} {\bibfnamefont {D.}~\bibnamefont {Blas}},
  \bibinfo {author} {\bibfnamefont {E.}~\bibnamefont {Barausse}}, \bibinfo
  {author} {\bibfnamefont {N.}~\bibnamefont {Cornish}}, \bibinfo {author}
  {\bibfnamefont {K.}~\bibnamefont {Yagi}},  and \bibinfo {author}
  {\bibfnamefont {N.}~\bibnamefont {Yunes}},\ }\href {\doibase
  10.1088/1361-6382/ac1a69} {\bibfield  {journal} {\bibinfo  {journal} {\emph
  {Class. Quant. Grav.}}\ }\textbf {\bibinfo {volume} {38}},\ \bibinfo {pages}
  {195003} (\bibinfo {year} {2021})},\ \Eprint
  {http://arxiv.org/abs/2104.04596} {arXiv:2104.04596 [gr-qc]} \BibitemShut
  {NoStop}%
\bibitem [{\citenamefont {Hansen}\ \emph {et~al.}(2015)\citenamefont {Hansen},
  \citenamefont {Yunes},\ and\ \citenamefont {Yagi}}]{Hansen:2014ewa}%
  \BibitemOpen
  \bibfield  {author} {\bibinfo {author} {\bibfnamefont {D.}~\bibnamefont
  {Hansen}}, \bibinfo {author} {\bibfnamefont {N.}~\bibnamefont {Yunes}},  and
  \bibinfo {author} {\bibfnamefont {K.}~\bibnamefont {Yagi}},\ }\href {\doibase
  10.1103/PhysRevD.91.082003} {\bibfield  {journal} {\bibinfo  {journal} {\emph
  {Phys. Rev. D}}\ }\textbf {\bibinfo {volume} {91}},\ \bibinfo {pages}
  {082003} (\bibinfo {year} {2015})},\ \Eprint {http://arxiv.org/abs/1412.4132}
  {arXiv:1412.4132 [gr-qc]} \BibitemShut {NoStop}%
\bibitem [{\citenamefont {Zhang}\ \emph
  {et~al.}(2020{\natexlab{a}})\citenamefont {Zhang}, \citenamefont {Zhao},
  \citenamefont {Wang}, \citenamefont {Wang}, \citenamefont {Yagi},
  \citenamefont {Yunes}, \citenamefont {Zhao},\ and\ \citenamefont
  {Zhu}}]{Zhang:2019iim}%
  \BibitemOpen
  \bibfield  {author} {\bibinfo {author} {\bibfnamefont {C.}~\bibnamefont
  {Zhang}}, \bibinfo {author} {\bibfnamefont {X.}~\bibnamefont {Zhao}},
  \bibinfo {author} {\bibfnamefont {A.}~\bibnamefont {Wang}}, \bibinfo {author}
  {\bibfnamefont {B.}~\bibnamefont {Wang}}, \bibinfo {author} {\bibfnamefont
  {K.}~\bibnamefont {Yagi}}, \bibinfo {author} {\bibfnamefont {N.}~\bibnamefont
  {Yunes}}, \bibinfo {author} {\bibfnamefont {W.}~\bibnamefont {Zhao}},  and
  \bibinfo {author} {\bibfnamefont {T.}~\bibnamefont {Zhu}},\ }\href {\doibase
  10.1103/PhysRevD.104.069905} {\bibfield  {journal} {\bibinfo  {journal}
  {\emph {Phys. Rev. D}}\ }\textbf {\bibinfo {volume} {101}},\ \bibinfo {pages}
  {044002} (\bibinfo {year} {2020}{\natexlab{a}})},\ \bibinfo {note} {[Erratum:
  Phys.Rev.D 104, 069905 (2021)]},\ \Eprint {http://arxiv.org/abs/1911.10278}
  {arXiv:1911.10278 [gr-qc]} \BibitemShut {NoStop}%
\bibitem [{\citenamefont {Schumacher}\ \emph {et~al.}(2023)\citenamefont
  {Schumacher}, \citenamefont {Perkins}, \citenamefont {Shaw}, \citenamefont
  {Yagi},\ and\ \citenamefont {Yunes}}]{Schumacher:2023cxh}%
  \BibitemOpen
  \bibfield  {author} {\bibinfo {author} {\bibfnamefont {K.}~\bibnamefont
  {Schumacher}}, \bibinfo {author} {\bibfnamefont {S.~E.}\ \bibnamefont
  {Perkins}}, \bibinfo {author} {\bibfnamefont {A.}~\bibnamefont {Shaw}},
  \bibinfo {author} {\bibfnamefont {K.}~\bibnamefont {Yagi}},  and \bibinfo
  {author} {\bibfnamefont {N.}~\bibnamefont {Yunes}},\ }\href {\doibase
  10.1103/PhysRevD.108.104053} {\bibfield  {journal} {\bibinfo  {journal}
  {\emph {Phys. Rev. D}}\ }\textbf {\bibinfo {volume} {108}},\ \bibinfo {pages}
  {104053} (\bibinfo {year} {2023})},\ \Eprint
  {http://arxiv.org/abs/2304.06801} {arXiv:2304.06801 [gr-qc]} \BibitemShut
  {NoStop}%
\bibitem [{\citenamefont {Eling}\ and\ \citenamefont
  {Jacobson}(2006)}]{Eling:2006ec}%
  \BibitemOpen
  \bibfield  {author} {\bibinfo {author} {\bibfnamefont {C.}~\bibnamefont
  {Eling}} and \bibinfo {author} {\bibfnamefont {T.}~\bibnamefont {Jacobson}},\
  }\href {\doibase 10.1088/0264-9381/23/18/009} {\bibfield  {journal} {\bibinfo
   {journal} {\emph {Class. Quant. Grav.}}\ }\textbf {\bibinfo {volume} {23}},\
  \bibinfo {pages} {5643} (\bibinfo {year} {2006})},\ \bibinfo {note}
  {[Erratum: Class.Quant.Grav. 27, 049802 (2010)]},\ \Eprint
  {http://arxiv.org/abs/gr-qc/0604088} {arXiv:gr-qc/0604088} \BibitemShut
  {NoStop}%
\bibitem [{\citenamefont {Foster}(2006)}]{Foster:2005fr}%
  \BibitemOpen
  \bibfield  {author} {\bibinfo {author} {\bibfnamefont {B.~Z.}\ \bibnamefont
  {Foster}},\ }\href {\doibase 10.1103/PhysRevD.73.024005} {\bibfield
  {journal} {\bibinfo  {journal} {\emph {Phys. Rev. D}}\ }\textbf {\bibinfo
  {volume} {73}},\ \bibinfo {pages} {024005} (\bibinfo {year} {2006})},\
  \Eprint {http://arxiv.org/abs/gr-qc/0509121} {arXiv:gr-qc/0509121}
  \BibitemShut {NoStop}%
\bibitem [{\citenamefont {Konoplya}\ and\ \citenamefont
  {Zhidenko}(2007)}]{Konoplya:2006rv}%
  \BibitemOpen
  \bibfield  {author} {\bibinfo {author} {\bibfnamefont {R.~A.}\ \bibnamefont
  {Konoplya}} and \bibinfo {author} {\bibfnamefont {A.}~\bibnamefont
  {Zhidenko}},\ }\href {\doibase 10.1016/j.physletb.2006.11.036} {\bibfield
  {journal} {\bibinfo  {journal} {\emph {Phys. Lett. B}}\ }\textbf {\bibinfo
  {volume} {644}},\ \bibinfo {pages} {186} (\bibinfo {year} {2007})},\ \Eprint
  {http://arxiv.org/abs/gr-qc/0605082} {arXiv:gr-qc/0605082} \BibitemShut
  {NoStop}%
\bibitem [{\citenamefont {Garfinkle}\ \emph {et~al.}(2007)\citenamefont
  {Garfinkle}, \citenamefont {Eling},\ and\ \citenamefont
  {Jacobson}}]{Garfinkle:2007bk}%
  \BibitemOpen
  \bibfield  {author} {\bibinfo {author} {\bibfnamefont {D.}~\bibnamefont
  {Garfinkle}}, \bibinfo {author} {\bibfnamefont {C.}~\bibnamefont {Eling}},
  and \bibinfo {author} {\bibfnamefont {T.}~\bibnamefont {Jacobson}},\ }\href
  {\doibase 10.1103/PhysRevD.76.024003} {\bibfield  {journal} {\bibinfo
  {journal} {\emph {Phys. Rev. D}}\ }\textbf {\bibinfo {volume} {76}},\
  \bibinfo {pages} {024003} (\bibinfo {year} {2007})},\ \Eprint
  {http://arxiv.org/abs/gr-qc/0703093} {arXiv:gr-qc/0703093} \BibitemShut
  {NoStop}%
\bibitem [{\citenamefont {Berglund}\ \emph {et~al.}(2012)\citenamefont
  {Berglund}, \citenamefont {Bhattacharyya},\ and\ \citenamefont
  {Mattingly}}]{Berglund:2012bu}%
  \BibitemOpen
  \bibfield  {author} {\bibinfo {author} {\bibfnamefont {P.}~\bibnamefont
  {Berglund}}, \bibinfo {author} {\bibfnamefont {J.}~\bibnamefont
  {Bhattacharyya}},  and \bibinfo {author} {\bibfnamefont {D.}~\bibnamefont
  {Mattingly}},\ }\href {\doibase 10.1103/PhysRevD.85.124019} {\bibfield
  {journal} {\bibinfo  {journal} {\emph {Phys. Rev. D}}\ }\textbf {\bibinfo
  {volume} {85}},\ \bibinfo {pages} {124019} (\bibinfo {year} {2012})},\
  \Eprint {http://arxiv.org/abs/1202.4497} {arXiv:1202.4497 [hep-th]}
  \BibitemShut {NoStop}%
\bibitem [{\citenamefont {Gao}\ and\ \citenamefont {Shen}(2013)}]{Gao:2013im}%
  \BibitemOpen
  \bibfield  {author} {\bibinfo {author} {\bibfnamefont {C.}~\bibnamefont
  {Gao}} and \bibinfo {author} {\bibfnamefont {Y.-G.}\ \bibnamefont {Shen}},\
  }\href {\doibase 10.1103/PhysRevD.88.103508} {\bibfield  {journal} {\bibinfo
  {journal} {\emph {Phys. Rev. D}}\ }\textbf {\bibinfo {volume} {88}},\
  \bibinfo {pages} {103508} (\bibinfo {year} {2013})},\ \Eprint
  {http://arxiv.org/abs/1301.7122} {arXiv:1301.7122 [gr-qc]} \BibitemShut
  {NoStop}%
\bibitem [{\citenamefont {Lin}\ \emph {et~al.}(2015)\citenamefont {Lin},
  \citenamefont {Goldoni}, \citenamefont {da~Silva},\ and\ \citenamefont
  {Wang}}]{Lin:2014eaa}%
  \BibitemOpen
  \bibfield  {author} {\bibinfo {author} {\bibfnamefont {K.}~\bibnamefont
  {Lin}}, \bibinfo {author} {\bibfnamefont {O.}~\bibnamefont {Goldoni}},
  \bibinfo {author} {\bibfnamefont {M.~F.}\ \bibnamefont {da~Silva}},  and
  \bibinfo {author} {\bibfnamefont {A.}~\bibnamefont {Wang}},\ }\href {\doibase
  10.1103/PhysRevD.91.024047} {\bibfield  {journal} {\bibinfo  {journal} {\emph
  {Phys. Rev. D}}\ }\textbf {\bibinfo {volume} {91}},\ \bibinfo {pages}
  {024047} (\bibinfo {year} {2015})},\ \Eprint {http://arxiv.org/abs/1410.6678}
  {arXiv:1410.6678 [gr-qc]} \BibitemShut {NoStop}%
\bibitem [{\citenamefont {Ding}\ \emph {et~al.}(2015)\citenamefont {Ding},
  \citenamefont {Wang},\ and\ \citenamefont {Wang}}]{Ding:2015kba}%
  \BibitemOpen
  \bibfield  {author} {\bibinfo {author} {\bibfnamefont {C.}~\bibnamefont
  {Ding}}, \bibinfo {author} {\bibfnamefont {A.}~\bibnamefont {Wang}},  and
  \bibinfo {author} {\bibfnamefont {X.}~\bibnamefont {Wang}},\ }\href {\doibase
  10.1103/PhysRevD.92.084055} {\bibfield  {journal} {\bibinfo  {journal} {\emph
  {Phys. Rev. D}}\ }\textbf {\bibinfo {volume} {92}},\ \bibinfo {pages}
  {084055} (\bibinfo {year} {2015})},\ \Eprint
  {http://arxiv.org/abs/1507.06618} {arXiv:1507.06618 [gr-qc]} \BibitemShut
  {NoStop}%
\bibitem [{\citenamefont {Ding}\ \emph {et~al.}(2016)\citenamefont {Ding},
  \citenamefont {Liu}, \citenamefont {Wang},\ and\ \citenamefont
  {Jing}}]{Ding:2016wcf}%
  \BibitemOpen
  \bibfield  {author} {\bibinfo {author} {\bibfnamefont {C.}~\bibnamefont
  {Ding}}, \bibinfo {author} {\bibfnamefont {C.}~\bibnamefont {Liu}}, \bibinfo
  {author} {\bibfnamefont {A.}~\bibnamefont {Wang}},  and \bibinfo {author}
  {\bibfnamefont {J.}~\bibnamefont {Jing}},\ }\href {\doibase
  10.1103/PhysRevD.94.124034} {\bibfield  {journal} {\bibinfo  {journal} {\emph
  {Phys. Rev. D}}\ }\textbf {\bibinfo {volume} {94}},\ \bibinfo {pages}
  {124034} (\bibinfo {year} {2016})},\ \Eprint
  {http://arxiv.org/abs/1608.00290} {arXiv:1608.00290 [gr-qc]} \BibitemShut
  {NoStop}%
\bibitem [{\citenamefont {Ding}(2017)}]{Ding:2017gfw}%
  \BibitemOpen
  \bibfield  {author} {\bibinfo {author} {\bibfnamefont {C.}~\bibnamefont
  {Ding}},\ }\href {\doibase 10.1103/PhysRevD.96.104021} {\bibfield  {journal}
  {\bibinfo  {journal} {\emph {Phys. Rev. D}}\ }\textbf {\bibinfo {volume}
  {96}},\ \bibinfo {pages} {104021} (\bibinfo {year} {2017})},\ \Eprint
  {http://arxiv.org/abs/1707.06747} {arXiv:1707.06747 [gr-qc]} \BibitemShut
  {NoStop}%
\bibitem [{\citenamefont {Lin}\ \emph {et~al.}(2018)\citenamefont {Lin},
  \citenamefont {Ho},\ and\ \citenamefont {Qian}}]{Lin:2017cmn}%
  \BibitemOpen
  \bibfield  {author} {\bibinfo {author} {\bibfnamefont {K.}~\bibnamefont
  {Lin}}, \bibinfo {author} {\bibfnamefont {F.-H.}\ \bibnamefont {Ho}},  and
  \bibinfo {author} {\bibfnamefont {W.-L.}\ \bibnamefont {Qian}},\ }\href
  {\doibase 10.1142/S0218271819500494} {\bibfield  {journal} {\bibinfo
  {journal} {\emph {Int. J. Mod. Phys. D}}\ }\textbf {\bibinfo {volume} {28}},\
  \bibinfo {pages} {1950049} (\bibinfo {year} {2018})},\ \Eprint
  {http://arxiv.org/abs/1704.06728} {arXiv:1704.06728 [gr-qc]} \BibitemShut
  {NoStop}%
\bibitem [{\citenamefont {Ding}\ and\ \citenamefont
  {Wang}(2019)}]{Ding:2018whp}%
  \BibitemOpen
  \bibfield  {author} {\bibinfo {author} {\bibfnamefont {C.}~\bibnamefont
  {Ding}} and \bibinfo {author} {\bibfnamefont {A.}~\bibnamefont {Wang}},\
  }\href {\doibase 10.1103/PhysRevD.99.124011} {\bibfield  {journal} {\bibinfo
  {journal} {\emph {Phys. Rev. D}}\ }\textbf {\bibinfo {volume} {99}},\
  \bibinfo {pages} {124011} (\bibinfo {year} {2019})},\ \Eprint
  {http://arxiv.org/abs/1811.05779} {arXiv:1811.05779 [gr-qc]} \BibitemShut
  {NoStop}%
\bibitem [{\citenamefont {Chan}\ \emph {et~al.}(2020)\citenamefont {Chan},
  \citenamefont {da~Silva},\ and\ \citenamefont
  {Satheeshkumar}}]{Chan:2019mdn}%
  \BibitemOpen
  \bibfield  {author} {\bibinfo {author} {\bibfnamefont {R.}~\bibnamefont
  {Chan}}, \bibinfo {author} {\bibfnamefont {M.~F.~A.}\ \bibnamefont
  {da~Silva}},  and \bibinfo {author} {\bibfnamefont {V.~H.}\ \bibnamefont
  {Satheeshkumar}},\ }\href {\doibase 10.1088/1475-7516/2020/05/025} {\bibfield
   {journal} {\bibinfo  {journal} {\emph {JCAP}}\ }\textbf {\bibinfo {volume}
  {05}},\ \bibinfo {pages} {025} (\bibinfo {year} {2020})},\ \Eprint
  {http://arxiv.org/abs/1912.12845} {arXiv:1912.12845 [gr-qc]} \BibitemShut
  {NoStop}%
\bibitem [{\citenamefont {Zhu}\ \emph {et~al.}(2019)\citenamefont {Zhu},
  \citenamefont {Wu}, \citenamefont {Jamil},\ and\ \citenamefont
  {Jusufi}}]{Zhu:2019ura}%
  \BibitemOpen
  \bibfield  {author} {\bibinfo {author} {\bibfnamefont {T.}~\bibnamefont
  {Zhu}}, \bibinfo {author} {\bibfnamefont {Q.}~\bibnamefont {Wu}}, \bibinfo
  {author} {\bibfnamefont {M.}~\bibnamefont {Jamil}},  and \bibinfo {author}
  {\bibfnamefont {K.}~\bibnamefont {Jusufi}},\ }\href {\doibase
  10.1103/PhysRevD.100.044055} {\bibfield  {journal} {\bibinfo  {journal}
  {\emph {Phys. Rev. D}}\ }\textbf {\bibinfo {volume} {100}},\ \bibinfo {pages}
  {044055} (\bibinfo {year} {2019})},\ \Eprint
  {http://arxiv.org/abs/1906.05673} {arXiv:1906.05673 [gr-qc]} \BibitemShut
  {NoStop}%
\bibitem [{\citenamefont {Chan}\ \emph {et~al.}(2021)\citenamefont {Chan},
  \citenamefont {Da~Silva},\ and\ \citenamefont
  {Satheeshkumar}}]{Chan:2020amr}%
  \BibitemOpen
  \bibfield  {author} {\bibinfo {author} {\bibfnamefont {R.}~\bibnamefont
  {Chan}}, \bibinfo {author} {\bibfnamefont {M.~F.~A.}\ \bibnamefont
  {Da~Silva}},  and \bibinfo {author} {\bibfnamefont {V.~H.}\ \bibnamefont
  {Satheeshkumar}},\ }\href {\doibase 10.1140/epjc/s10052-021-09120-w}
  {\bibfield  {journal} {\bibinfo  {journal} {\emph {Eur. Phys. J. C}}\
  }\textbf {\bibinfo {volume} {81}},\ \bibinfo {pages} {317} (\bibinfo {year}
  {2021})},\ \Eprint {http://arxiv.org/abs/2003.00227} {arXiv:2003.00227
  [gr-qc]} \BibitemShut {NoStop}%
\bibitem [{\citenamefont {Khodadi}\ and\ \citenamefont
  {Saridakis}(2021)}]{Khodadi:2020gns}%
  \BibitemOpen
  \bibfield  {author} {\bibinfo {author} {\bibfnamefont {M.}~\bibnamefont
  {Khodadi}} and \bibinfo {author} {\bibfnamefont {E.~N.}\ \bibnamefont
  {Saridakis}},\ }\href {\doibase 10.1016/j.dark.2021.100835} {\bibfield
  {journal} {\bibinfo  {journal} {\emph {Phys. Dark Univ.}}\ }\textbf {\bibinfo
  {volume} {32}},\ \bibinfo {pages} {100835} (\bibinfo {year} {2021})},\
  \Eprint {http://arxiv.org/abs/2012.05186} {arXiv:2012.05186 [gr-qc]}
  \BibitemShut {NoStop}%
\bibitem [{\citenamefont {Zhang}\ \emph
  {et~al.}(2020{\natexlab{b}})\citenamefont {Zhang}, \citenamefont {Zhao},
  \citenamefont {Lin}, \citenamefont {Zhang}, \citenamefont {Zhao},\ and\
  \citenamefont {Wang}}]{Zhang:2020too}%
  \BibitemOpen
  \bibfield  {author} {\bibinfo {author} {\bibfnamefont {C.}~\bibnamefont
  {Zhang}}, \bibinfo {author} {\bibfnamefont {X.}~\bibnamefont {Zhao}},
  \bibinfo {author} {\bibfnamefont {K.}~\bibnamefont {Lin}}, \bibinfo {author}
  {\bibfnamefont {S.}~\bibnamefont {Zhang}}, \bibinfo {author} {\bibfnamefont
  {W.}~\bibnamefont {Zhao}},  and \bibinfo {author} {\bibfnamefont
  {A.}~\bibnamefont {Wang}},\ }\href {\doibase 10.1103/PhysRevD.102.064043}
  {\bibfield  {journal} {\bibinfo  {journal} {\emph {Phys. Rev. D}}\ }\textbf
  {\bibinfo {volume} {102}},\ \bibinfo {pages} {064043} (\bibinfo {year}
  {2020}{\natexlab{b}})},\ \Eprint {http://arxiv.org/abs/2004.06155}
  {arXiv:2004.06155 [gr-qc]} \BibitemShut {NoStop}%
\bibitem [{\citenamefont {Oost}\ \emph {et~al.}(2021)\citenamefont {Oost},
  \citenamefont {Mukohyama},\ and\ \citenamefont {Wang}}]{Oost:2021tqi}%
  \BibitemOpen
  \bibfield  {author} {\bibinfo {author} {\bibfnamefont {J.}~\bibnamefont
  {Oost}}, \bibinfo {author} {\bibfnamefont {S.}~\bibnamefont {Mukohyama}},
  and \bibinfo {author} {\bibfnamefont {A.}~\bibnamefont {Wang}},\ }\href
  {\doibase 10.3390/universe7080272} {\bibfield  {journal} {\bibinfo  {journal}
  {\emph {Universe}}\ }\textbf {\bibinfo {volume} {7}},\ \bibinfo {pages} {272}
  (\bibinfo {year} {2021})},\ \Eprint {http://arxiv.org/abs/2106.09044}
  {arXiv:2106.09044 [gr-qc]} \BibitemShut {NoStop}%
\bibitem [{\citenamefont {Mazza}\ and\ \citenamefont
  {Liberati}(2023)}]{Mazza:2023iwv}%
  \BibitemOpen
  \bibfield  {author} {\bibinfo {author} {\bibfnamefont {J.}~\bibnamefont
  {Mazza}} and \bibinfo {author} {\bibfnamefont {S.}~\bibnamefont {Liberati}},\
  }\href {\doibase 10.1007/JHEP03(2023)199} {\bibfield  {journal} {\bibinfo
  {journal} {\emph {JHEP}}\ }\textbf {\bibinfo {volume} {03}},\ \bibinfo
  {pages} {199} (\bibinfo {year} {2023})},\ \Eprint
  {http://arxiv.org/abs/2301.04697} {arXiv:2301.04697 [gr-qc]} \BibitemShut
  {NoStop}%
\bibitem [{\citenamefont {Blas}\ and\ \citenamefont
  {Sibiryakov}(2011)}]{Blas:2011ni}%
  \BibitemOpen
  \bibfield  {author} {\bibinfo {author} {\bibfnamefont {D.}~\bibnamefont
  {Blas}} and \bibinfo {author} {\bibfnamefont {S.}~\bibnamefont
  {Sibiryakov}},\ }\href {\doibase 10.1103/PhysRevD.84.124043} {\bibfield
  {journal} {\bibinfo  {journal} {\emph {Phys. Rev. D}}\ }\textbf {\bibinfo
  {volume} {84}},\ \bibinfo {pages} {124043} (\bibinfo {year} {2011})},\
  \Eprint {http://arxiv.org/abs/1110.2195} {arXiv:1110.2195 [hep-th]}
  \BibitemShut {NoStop}%
\bibitem [{\citenamefont {Regge}\ and\ \citenamefont
  {Wheeler}(1957)}]{Regge:1957td}%
  \BibitemOpen
  \bibfield  {author} {\bibinfo {author} {\bibfnamefont {T.}~\bibnamefont
  {Regge}} and \bibinfo {author} {\bibfnamefont {J.~A.}\ \bibnamefont
  {Wheeler}},\ }\href {\doibase 10.1103/PhysRev.108.1063} {\bibfield  {journal}
  {\bibinfo  {journal} {\emph {Phys. Rev.}}\ }\textbf {\bibinfo {volume}
  {108}},\ \bibinfo {pages} {1063} (\bibinfo {year} {1957})}\BibitemShut
  {NoStop}%
\bibitem [{\citenamefont {Zerilli}(1970)}]{Zerilli:1970se}%
  \BibitemOpen
  \bibfield  {author} {\bibinfo {author} {\bibfnamefont {F.~J.}\ \bibnamefont
  {Zerilli}},\ }\href {\doibase 10.1103/PhysRevLett.24.737} {\bibfield
  {journal} {\bibinfo  {journal} {\emph {Phys. Rev. Lett.}}\ }\textbf {\bibinfo
  {volume} {24}},\ \bibinfo {pages} {737} (\bibinfo {year} {1970})}\BibitemShut
  {NoStop}%
\bibitem [{\citenamefont {Horndeski}(1974)}]{Horndeski:1974wa}%
  \BibitemOpen
  \bibfield  {author} {\bibinfo {author} {\bibfnamefont {G.~W.}\ \bibnamefont
  {Horndeski}},\ }\href {\doibase 10.1007/BF01807638} {\bibfield  {journal}
  {\bibinfo  {journal} {\emph {Int. J. Theor. Phys.}}\ }\textbf {\bibinfo
  {volume} {10}},\ \bibinfo {pages} {363} (\bibinfo {year} {1974})}\BibitemShut
  {NoStop}%
\bibitem [{\citenamefont {Kobayashi}\ \emph {et~al.}(2012)\citenamefont
  {Kobayashi}, \citenamefont {Motohashi},\ and\ \citenamefont
  {Suyama}}]{Kobayashi:2012kh}%
  \BibitemOpen
  \bibfield  {author} {\bibinfo {author} {\bibfnamefont {T.}~\bibnamefont
  {Kobayashi}}, \bibinfo {author} {\bibfnamefont {H.}~\bibnamefont
  {Motohashi}},  and \bibinfo {author} {\bibfnamefont {T.}~\bibnamefont
  {Suyama}},\ }\href {\doibase 10.1103/PhysRevD.85.084025} {\bibfield
  {journal} {\bibinfo  {journal} {\emph {Phys. Rev. D}}\ }\textbf {\bibinfo
  {volume} {85}},\ \bibinfo {pages} {084025} (\bibinfo {year} {2012})},\
  \bibinfo {note} {[Erratum: Phys.Rev.D 96, 109903 (2017)]},\ \Eprint
  {http://arxiv.org/abs/1202.4893} {arXiv:1202.4893 [gr-qc]} \BibitemShut
  {NoStop}%
\bibitem [{\citenamefont {Kobayashi}\ \emph {et~al.}(2014)\citenamefont
  {Kobayashi}, \citenamefont {Motohashi},\ and\ \citenamefont
  {Suyama}}]{Kobayashi:2014wsa}%
  \BibitemOpen
  \bibfield  {author} {\bibinfo {author} {\bibfnamefont {T.}~\bibnamefont
  {Kobayashi}}, \bibinfo {author} {\bibfnamefont {H.}~\bibnamefont
  {Motohashi}},  and \bibinfo {author} {\bibfnamefont {T.}~\bibnamefont
  {Suyama}},\ }\href {\doibase 10.1103/PhysRevD.89.084042} {\bibfield
  {journal} {\bibinfo  {journal} {\emph {Phys. Rev. D}}\ }\textbf {\bibinfo
  {volume} {89}},\ \bibinfo {pages} {084042} (\bibinfo {year} {2014})},\
  \Eprint {http://arxiv.org/abs/1402.6740} {arXiv:1402.6740 [gr-qc]}
  \BibitemShut {NoStop}%
\bibitem [{\citenamefont {Kase}\ and\ \citenamefont
  {Tsujikawa}(2022)}]{Kase:2021mix}%
  \BibitemOpen
  \bibfield  {author} {\bibinfo {author} {\bibfnamefont {R.}~\bibnamefont
  {Kase}} and \bibinfo {author} {\bibfnamefont {S.}~\bibnamefont {Tsujikawa}},\
  }\href {\doibase 10.1103/PhysRevD.105.024059} {\bibfield  {journal} {\bibinfo
   {journal} {\emph {Phys. Rev. D}}\ }\textbf {\bibinfo {volume} {105}},\
  \bibinfo {pages} {024059} (\bibinfo {year} {2022})},\ \Eprint
  {http://arxiv.org/abs/2110.12728} {arXiv:2110.12728 [gr-qc]} \BibitemShut
  {NoStop}%
\bibitem [{\citenamefont {Minamitsuji}\ \emph
  {et~al.}(2022{\natexlab{a}})\citenamefont {Minamitsuji}, \citenamefont
  {Takahashi},\ and\ \citenamefont {Tsujikawa}}]{Minamitsuji:2022mlv}%
  \BibitemOpen
  \bibfield  {author} {\bibinfo {author} {\bibfnamefont {M.}~\bibnamefont
  {Minamitsuji}}, \bibinfo {author} {\bibfnamefont {K.}~\bibnamefont
  {Takahashi}},  and \bibinfo {author} {\bibfnamefont {S.}~\bibnamefont
  {Tsujikawa}},\ }\href {\doibase 10.1103/PhysRevD.105.104001} {\bibfield
  {journal} {\bibinfo  {journal} {\emph {Phys. Rev. D}}\ }\textbf {\bibinfo
  {volume} {105}},\ \bibinfo {pages} {104001} (\bibinfo {year}
  {2022}{\natexlab{a}})},\ \Eprint {http://arxiv.org/abs/2201.09687}
  {arXiv:2201.09687 [gr-qc]} \BibitemShut {NoStop}%
\bibitem [{\citenamefont {Tsujikawa}(2022)}]{Tsujikawa:2022lww}%
  \BibitemOpen
  \bibfield  {author} {\bibinfo {author} {\bibfnamefont {S.}~\bibnamefont
  {Tsujikawa}},\ }\href {\doibase 10.1016/j.physletb.2022.137329} {\bibfield
  {journal} {\bibinfo  {journal} {\emph {Phys. Lett. B}}\ }\textbf {\bibinfo
  {volume} {833}},\ \bibinfo {pages} {137329} (\bibinfo {year} {2022})},\
  \Eprint {http://arxiv.org/abs/2205.09932} {arXiv:2205.09932 [gr-qc]}
  \BibitemShut {NoStop}%
\bibitem [{\citenamefont {Kase}\ and\ \citenamefont
  {Tsujikawa}(2023)}]{Kase:2023kvq}%
  \BibitemOpen
  \bibfield  {author} {\bibinfo {author} {\bibfnamefont {R.}~\bibnamefont
  {Kase}} and \bibinfo {author} {\bibfnamefont {S.}~\bibnamefont {Tsujikawa}},\
  }\href {\doibase 10.1103/PhysRevD.107.104045} {\bibfield  {journal} {\bibinfo
   {journal} {\emph {Phys. Rev. D}}\ }\textbf {\bibinfo {volume} {107}},\
  \bibinfo {pages} {104045} (\bibinfo {year} {2023})},\ \Eprint
  {http://arxiv.org/abs/2301.10362} {arXiv:2301.10362 [gr-qc]} \BibitemShut
  {NoStop}%
\bibitem [{\citenamefont {Minamitsuji}\ \emph
  {et~al.}(2022{\natexlab{b}})\citenamefont {Minamitsuji}, \citenamefont
  {Takahashi},\ and\ \citenamefont {Tsujikawa}}]{Minamitsuji:2022vbi}%
  \BibitemOpen
  \bibfield  {author} {\bibinfo {author} {\bibfnamefont {M.}~\bibnamefont
  {Minamitsuji}}, \bibinfo {author} {\bibfnamefont {K.}~\bibnamefont
  {Takahashi}},  and \bibinfo {author} {\bibfnamefont {S.}~\bibnamefont
  {Tsujikawa}},\ }\href {\doibase 10.1103/PhysRevD.106.044003} {\bibfield
  {journal} {\bibinfo  {journal} {\emph {Phys. Rev. D}}\ }\textbf {\bibinfo
  {volume} {106}},\ \bibinfo {pages} {044003} (\bibinfo {year}
  {2022}{\natexlab{b}})},\ \Eprint {http://arxiv.org/abs/2204.13837}
  {arXiv:2204.13837 [gr-qc]} \BibitemShut {NoStop}%
\bibitem [{\citenamefont {Tsujikawa}\ \emph {et~al.}(2021)\citenamefont
  {Tsujikawa}, \citenamefont {Zhang}, \citenamefont {Zhao},\ and\ \citenamefont
  {Wang}}]{Tsujikawa:2021typ}%
  \BibitemOpen
  \bibfield  {author} {\bibinfo {author} {\bibfnamefont {S.}~\bibnamefont
  {Tsujikawa}}, \bibinfo {author} {\bibfnamefont {C.}~\bibnamefont {Zhang}},
  \bibinfo {author} {\bibfnamefont {X.}~\bibnamefont {Zhao}},  and \bibinfo
  {author} {\bibfnamefont {A.}~\bibnamefont {Wang}},\ }\href {\doibase
  10.1103/PhysRevD.104.064024} {\bibfield  {journal} {\bibinfo  {journal}
  {\emph {Phys. Rev. D}}\ }\textbf {\bibinfo {volume} {104}},\ \bibinfo {pages}
  {064024} (\bibinfo {year} {2021})},\ \Eprint
  {http://arxiv.org/abs/2107.08061} {arXiv:2107.08061 [gr-qc]} \BibitemShut
  {NoStop}%
\bibitem [{\citenamefont {Blas}\ \emph {et~al.}(2011)\citenamefont {Blas},
  \citenamefont {Pujolas},\ and\ \citenamefont {Sibiryakov}}]{Blas:2010hb}%
  \BibitemOpen
  \bibfield  {author} {\bibinfo {author} {\bibfnamefont {D.}~\bibnamefont
  {Blas}}, \bibinfo {author} {\bibfnamefont {O.}~\bibnamefont {Pujolas}},  and
  \bibinfo {author} {\bibfnamefont {S.}~\bibnamefont {Sibiryakov}},\ }\href
  {\doibase 10.1007/JHEP04(2011)018} {\bibfield  {journal} {\bibinfo  {journal}
  {\emph {JHEP}}\ }\textbf {\bibinfo {volume} {04}},\ \bibinfo {pages} {018}
  (\bibinfo {year} {2011})},\ \Eprint {http://arxiv.org/abs/1007.3503}
  {arXiv:1007.3503 [hep-th]} \BibitemShut {NoStop}%
\bibitem [{\citenamefont {Jacobson}(2010)}]{Jacobson:2010mx}%
  \BibitemOpen
  \bibfield  {author} {\bibinfo {author} {\bibfnamefont {T.}~\bibnamefont
  {Jacobson}},\ }\href {\doibase 10.1103/PhysRevD.81.101502} {\bibfield
  {journal} {\bibinfo  {journal} {\emph {Phys. Rev. D}}\ }\textbf {\bibinfo
  {volume} {81}},\ \bibinfo {pages} {101502} (\bibinfo {year} {2010})},\
  \bibinfo {note} {[Erratum: Phys.Rev.D 82, 129901 (2010)]},\ \Eprint
  {http://arxiv.org/abs/1001.4823} {arXiv:1001.4823 [hep-th]} \BibitemShut
  {NoStop}%
\bibitem [{\citenamefont {Blas}\ \emph {et~al.}(2010)\citenamefont {Blas},
  \citenamefont {Pujolas},\ and\ \citenamefont {Sibiryakov}}]{Blas:2009qj}%
  \BibitemOpen
  \bibfield  {author} {\bibinfo {author} {\bibfnamefont {D.}~\bibnamefont
  {Blas}}, \bibinfo {author} {\bibfnamefont {O.}~\bibnamefont {Pujolas}},  and
  \bibinfo {author} {\bibfnamefont {S.}~\bibnamefont {Sibiryakov}},\ }\href
  {\doibase 10.1103/PhysRevLett.104.181302} {\bibfield  {journal} {\bibinfo
  {journal} {\emph {Phys. Rev. Lett.}}\ }\textbf {\bibinfo {volume} {104}},\
  \bibinfo {pages} {181302} (\bibinfo {year} {2010})},\ \Eprint
  {http://arxiv.org/abs/0909.3525} {arXiv:0909.3525 [hep-th]} \BibitemShut
  {NoStop}%
\bibitem [{\citenamefont {Horava}(2009)}]{Horava:2009uw}%
  \BibitemOpen
  \bibfield  {author} {\bibinfo {author} {\bibfnamefont {P.}~\bibnamefont
  {Horava}},\ }\href {\doibase 10.1103/PhysRevD.79.084008} {\bibfield
  {journal} {\bibinfo  {journal} {\emph {Phys. Rev. D}}\ }\textbf {\bibinfo
  {volume} {79}},\ \bibinfo {pages} {084008} (\bibinfo {year} {2009})},\
  \Eprint {http://arxiv.org/abs/0901.3775} {arXiv:0901.3775 [hep-th]}
  \BibitemShut {NoStop}%
\bibitem [{\citenamefont {Lin}\ \emph {et~al.}(2017)\citenamefont {Lin},
  \citenamefont {Mukohyama}, \citenamefont {Wang},\ and\ \citenamefont
  {Zhu}}]{Lin:2017jvc}%
  \BibitemOpen
  \bibfield  {author} {\bibinfo {author} {\bibfnamefont {K.}~\bibnamefont
  {Lin}}, \bibinfo {author} {\bibfnamefont {S.}~\bibnamefont {Mukohyama}},
  \bibinfo {author} {\bibfnamefont {A.}~\bibnamefont {Wang}},  and \bibinfo
  {author} {\bibfnamefont {T.}~\bibnamefont {Zhu}},\ }\href {\doibase
  10.1103/PhysRevD.95.124053} {\bibfield  {journal} {\bibinfo  {journal} {\emph
  {Phys. Rev. D}}\ }\textbf {\bibinfo {volume} {95}},\ \bibinfo {pages}
  {124053} (\bibinfo {year} {2017})},\ \Eprint
  {http://arxiv.org/abs/1704.02990} {arXiv:1704.02990 [gr-qc]} \BibitemShut
  {NoStop}%
\bibitem [{\citenamefont {Wang}(2017)}]{Wang:2017brl}%
  \BibitemOpen
  \bibfield  {author} {\bibinfo {author} {\bibfnamefont {A.}~\bibnamefont
  {Wang}},\ }\href {\doibase 10.1142/S0218271817300142} {\bibfield  {journal}
  {\bibinfo  {journal} {\emph {Int. J. Mod. Phys. D}}\ }\textbf {\bibinfo
  {volume} {26}},\ \bibinfo {pages} {1730014} (\bibinfo {year} {2017})},\
  \Eprint {http://arxiv.org/abs/1701.06087} {arXiv:1701.06087 [gr-qc]}
  \BibitemShut {NoStop}%
\bibitem [{\citenamefont {Foster}(2005)}]{Foster:2005ec}%
  \BibitemOpen
  \bibfield  {author} {\bibinfo {author} {\bibfnamefont {B.~Z.}\ \bibnamefont
  {Foster}},\ }\href {\doibase 10.1103/PhysRevD.72.044017} {\bibfield
  {journal} {\bibinfo  {journal} {\emph {Phys. Rev. D}}\ }\textbf {\bibinfo
  {volume} {72}},\ \bibinfo {pages} {044017} (\bibinfo {year} {2005})},\
  \Eprint {http://arxiv.org/abs/gr-qc/0502066} {arXiv:gr-qc/0502066}
  \BibitemShut {NoStop}%
\bibitem [{\citenamefont {Jacobson}(2007)}]{Jacobson:2007veq}%
  \BibitemOpen
  \bibfield  {author} {\bibinfo {author} {\bibfnamefont {T.}~\bibnamefont
  {Jacobson}},\ }\href {\doibase 10.22323/1.043.0020} {\bibfield  {journal}
  {\bibinfo  {journal} {\emph {PoS}}\ }\textbf {\bibinfo {volume} {QG-PH}},\
  \bibinfo {pages} {020} (\bibinfo {year} {2007})},\ \Eprint
  {http://arxiv.org/abs/0801.1547} {arXiv:0801.1547 [gr-qc]} \BibitemShut
  {NoStop}%
\bibitem [{\citenamefont {Bekenstein}(1993)}]{Bekenstein:1992pj}%
  \BibitemOpen
  \bibfield  {author} {\bibinfo {author} {\bibfnamefont {J.~D.}\ \bibnamefont
  {Bekenstein}},\ }\href {\doibase 10.1103/PhysRevD.48.3641} {\bibfield
  {journal} {\bibinfo  {journal} {\emph {Phys. Rev. D}}\ }\textbf {\bibinfo
  {volume} {48}},\ \bibinfo {pages} {3641} (\bibinfo {year} {1993})},\ \Eprint
  {http://arxiv.org/abs/gr-qc/9211017} {arXiv:gr-qc/9211017} \BibitemShut
  {NoStop}%
\bibitem [{\citenamefont {Dom\`enech}\ \emph {et~al.}(2018)\citenamefont
  {Dom\`enech}, \citenamefont {Mukohyama}, \citenamefont {Namba},\ and\
  \citenamefont {Papadopoulos}}]{Domenech:2018vqj}%
  \BibitemOpen
  \bibfield  {author} {\bibinfo {author} {\bibfnamefont {G.}~\bibnamefont
  {Dom\`enech}}, \bibinfo {author} {\bibfnamefont {S.}~\bibnamefont
  {Mukohyama}}, \bibinfo {author} {\bibfnamefont {R.}~\bibnamefont {Namba}},
  and \bibinfo {author} {\bibfnamefont {V.}~\bibnamefont {Papadopoulos}},\
  }\href {\doibase 10.1103/PhysRevD.98.064037} {\bibfield  {journal} {\bibinfo
  {journal} {\emph {Phys. Rev. D}}\ }\textbf {\bibinfo {volume} {98}},\
  \bibinfo {pages} {064037} (\bibinfo {year} {2018})},\ \Eprint
  {http://arxiv.org/abs/1807.06048} {arXiv:1807.06048 [gr-qc]} \BibitemShut
  {NoStop}%
\bibitem [{\citenamefont {De~Felice}\ and\ \citenamefont
  {Mukohyama}(2016)}]{DeFelice:2015hla}%
  \BibitemOpen
  \bibfield  {author} {\bibinfo {author} {\bibfnamefont {A.}~\bibnamefont
  {De~Felice}} and \bibinfo {author} {\bibfnamefont {S.}~\bibnamefont
  {Mukohyama}},\ }\href {\doibase 10.1016/j.physletb.2015.11.050} {\bibfield
  {journal} {\bibinfo  {journal} {\emph {Phys. Lett. B}}\ }\textbf {\bibinfo
  {volume} {752}},\ \bibinfo {pages} {302} (\bibinfo {year} {2016})},\ \Eprint
  {http://arxiv.org/abs/1506.01594} {arXiv:1506.01594 [hep-th]} \BibitemShut
  {NoStop}%
\bibitem [{\citenamefont {G\"umr\"uk\c{c}\"uo\u{g}lu}\ \emph
  {et~al.}(2016)\citenamefont {G\"umr\"uk\c{c}\"uo\u{g}lu}, \citenamefont
  {Mukohyama},\ and\ \citenamefont {Sotiriou}}]{Gumrukcuoglu:2016jbh}%
  \BibitemOpen
  \bibfield  {author} {\bibinfo {author} {\bibfnamefont {A.~E.}\ \bibnamefont
  {G\"umr\"uk\c{c}\"uo\u{g}lu}}, \bibinfo {author} {\bibfnamefont
  {S.}~\bibnamefont {Mukohyama}},  and \bibinfo {author} {\bibfnamefont
  {T.~P.}\ \bibnamefont {Sotiriou}},\ }\href {\doibase
  10.1103/PhysRevD.94.064001} {\bibfield  {journal} {\bibinfo  {journal} {\emph
  {Phys. Rev. D}}\ }\textbf {\bibinfo {volume} {94}},\ \bibinfo {pages}
  {064001} (\bibinfo {year} {2016})},\ \Eprint
  {http://arxiv.org/abs/1606.00618} {arXiv:1606.00618 [hep-th]} \BibitemShut
  {NoStop}%
\bibitem [{\citenamefont {Kubota}\ \emph {et~al.}(2023)\citenamefont {Kubota},
  \citenamefont {Arai},\ and\ \citenamefont {Mukohyama}}]{Kubota:2022lbn}%
  \BibitemOpen
  \bibfield  {author} {\bibinfo {author} {\bibfnamefont {K.-i.}\ \bibnamefont
  {Kubota}}, \bibinfo {author} {\bibfnamefont {S.}~\bibnamefont {Arai}},  and
  \bibinfo {author} {\bibfnamefont {S.}~\bibnamefont {Mukohyama}},\ }\href
  {\doibase 10.1103/PhysRevD.107.064002} {\bibfield  {journal} {\bibinfo
  {journal} {\emph {Phys. Rev. D}}\ }\textbf {\bibinfo {volume} {107}},\
  \bibinfo {pages} {064002} (\bibinfo {year} {2023})},\ \Eprint
  {http://arxiv.org/abs/2209.00795} {arXiv:2209.00795 [gr-qc]} \BibitemShut
  {NoStop}%
\bibitem [{\citenamefont {Mukohyama}(2002)}]{Mukohyama:2002vq}%
  \BibitemOpen
  \bibfield  {author} {\bibinfo {author} {\bibfnamefont {S.}~\bibnamefont
  {Mukohyama}},\ }\href {\doibase 10.1103/PhysRevD.66.123512} {\bibfield
  {journal} {\bibinfo  {journal} {\emph {Phys. Rev. D}}\ }\textbf {\bibinfo
  {volume} {66}},\ \bibinfo {pages} {123512} (\bibinfo {year} {2002})},\
  \Eprint {http://arxiv.org/abs/hep-th/0208094} {arXiv:hep-th/0208094}
  \BibitemShut {NoStop}%
\bibitem [{\citenamefont {Barausse}\ \emph {et~al.}(2011)\citenamefont
  {Barausse}, \citenamefont {Jacobson},\ and\ \citenamefont
  {Sotiriou}}]{Barausse:2011pu}%
  \BibitemOpen
  \bibfield  {author} {\bibinfo {author} {\bibfnamefont {E.}~\bibnamefont
  {Barausse}}, \bibinfo {author} {\bibfnamefont {T.}~\bibnamefont {Jacobson}},
  and \bibinfo {author} {\bibfnamefont {T.~P.}\ \bibnamefont {Sotiriou}},\
  }\href {\doibase 10.1103/PhysRevD.83.124043} {\bibfield  {journal} {\bibinfo
  {journal} {\emph {Phys. Rev. D}}\ }\textbf {\bibinfo {volume} {83}},\
  \bibinfo {pages} {124043} (\bibinfo {year} {2011})},\ \Eprint
  {http://arxiv.org/abs/1104.2889} {arXiv:1104.2889 [gr-qc]} \BibitemShut
  {NoStop}%
\bibitem [{\citenamefont {Zhang}\ \emph {et~al.}(2023)\citenamefont {Zhang},
  \citenamefont {Wang},\ and\ \citenamefont {Zhu}}]{Zhang:2022fbz}%
  \BibitemOpen
  \bibfield  {author} {\bibinfo {author} {\bibfnamefont {C.}~\bibnamefont
  {Zhang}}, \bibinfo {author} {\bibfnamefont {A.}~\bibnamefont {Wang}},  and
  \bibinfo {author} {\bibfnamefont {T.}~\bibnamefont {Zhu}},\ }\href {\doibase
  10.1140/epjc/s10052-023-11998-7} {\bibfield  {journal} {\bibinfo  {journal}
  {\emph {Eur. Phys. J. C}}\ }\textbf {\bibinfo {volume} {83}},\ \bibinfo
  {pages} {841} (\bibinfo {year} {2023})},\ \Eprint
  {http://arxiv.org/abs/2209.04735} {arXiv:2209.04735 [gr-qc]} \BibitemShut
  {NoStop}%
\end{thebibliography}%

\end{document}